\documentclass[twocolumn]{aastex631}

\usepackage{natbib}
\usepackage{subfigure}
\usepackage{mathrsfs}
\usepackage{booktabs}
\usepackage{amsmath}
\usepackage{bm}
\usepackage{morefloats}

\newcommand{\aox}{$\alpha_{\rm OX}$}
\newcommand{\ltkf}{$L_{\rm 2500~\textup{\AA}}$}
\newcommand{\ftkf}{$f_{\rm 2500~\textup{\AA}}$}
\newcommand{\ltkev}{$L_{\rm 2~keV}$}
\newcommand{\ftkev}{$f_{\rm 2~keV}$}
\newcommand{\xray}{\hbox{X-ray}}

\newcommand{\xmm}{\hbox{XMM-Newton}}

\newcommand{\galex}{{GALEX}}

\graphicspath{{./}{figures/}}

\begin{document}

\title{Photometric Selection of \hbox{type 1} Quasars in 
the XMM-LSS Field with Machine Learning and the Disk-Corona Connection
}

\author[0000-0002-9335-9455]{Jian Huang}
\affiliation{School of Astronomy and Space Science, Nanjing University, Nanjing, Jiangsu 210093, People's Republic of China}
\affiliation{Key Laboratory of Modern Astronomy and Astrophysics (Nanjing University), Ministry of Education, Nanjing 210093, People's Republic of China}

\author{Bin Luo}
\affiliation{School of Astronomy and Space Science, Nanjing University, Nanjing, Jiangsu 210093, People's Republic of China}
\affiliation{Key Laboratory of Modern Astronomy and Astrophysics (Nanjing University), Ministry of Education, Nanjing 210093, People's Republic of China}

\author{W. N. Brandt}
\affiliation{Department of Astronomy \& Astrophysics, 525 Davey Lab,
	The Pennsylvania State University, University Park, PA 16802, USA}
\affiliation{Institute for Gravitation and the Cosmos,
	The Pennsylvania State University, University Park, PA 16802, USA}
\affiliation{Department of Physics, 104 Davey Lab, The Pennsylvania State University, University Park, PA 16802, USA}

\author{Ying Chen}
\affiliation{School of Astronomy and Space Science, Nanjing University, Nanjing, Jiangsu 210093, People's Republic of China}
\affiliation{Key Laboratory of Modern Astronomy and Astrophysics (Nanjing University), Ministry of Education, Nanjing 210093, People's Republic of China}

\author{Qingling Ni}
\affiliation{Max-Planck-Institut f\"{u}r extraterrestrische Physik (MPE), Gie{\ss}enbachstra{\ss}e 1, D-85748 Garching bei M\"unchen, Germany}

\author{Yongquan Xue}
\affiliation{CAS Key Laboratory for Research in Galaxies and Cosmology, Department of Astronomy, University of Science and Technology of China, Hefei 230026, China}
\affiliation{School of Astronomy and Space Sciences, University of Science and Technology of China, Hefei 230026, China}


\author{Zijian Zhang}
\affiliation{Department of Astronomy, School of Physics, Peking University, Beijing 100871, People's Republic of China}
\affiliation{School of Astronomy and Space Science, Nanjing University, Nanjing, Jiangsu 210093, People's Republic of China}

\begin{abstract}
We present photometric selection of \hbox{type 1} quasars in the $\approx5.3~{\rm deg}^{2}$ \hbox{XMM-Large} Scale Structure (\hbox{XMM-LSS}) survey field with machine learning.
We constructed our training and \hbox{blind-test} samples using spectroscopically identified SDSS quasars, galaxies, and stars.
We utilized the \hbox{XGBoost} machine learning method to select a total of 1\,591 quasars.
We assessed the classification performance based on the \hbox{blind-test} sample, and the outcome was favorable, demonstrating high reliability ($\approx99.9\%$) and good completeness ($\approx87.5\%$).
We used \hbox{XGBoost} to estimate photometric redshifts of our selected quasars.
The estimated photometric redshifts span a range from 0.41 to 3.75.
The outlier fraction of these photometric redshift estimates is $\approx17\%$ and the normalized median absolute deviation ($\sigma_{\rm NMAD}$) is $\approx0.07$.
To study the quasar \hbox{disk-corona} connection, we constructed a subsample of 1\,016 quasars with HSC $i<22.5$ after excluding \hbox{radio-loud} and potentially \hbox{\xray-absorbed} quasars.
The relation between the \hbox{optical-to-X-ray} \hbox{power-law} slope parameter (\aox) and the  $2500~\textup{\AA}$ monochromatic luminosity ($L_{2500~\textup{\AA}}$) for this subsample is $\alpha_{\rm OX}=(-0.156\pm0.007)~{\rm log}~{L_{\rm 2500~\textup{\AA}}}+(3.175\pm0.211)$ with a dispersion of 0.159.
We found this correlation in good agreement with the correlations in previous studies.
We explored several factors which may bias the \hbox{\aox--\ltkf} relation and found that their effects are not significant.
We discussed possible evolution of the \hbox{\aox--\ltkf} relation with respect to \ltkf\ or redshift. 
\\
Unified Astronomy Thesaurus concepts: Quasars (1319); Radio quiet quasars (1354); Supermassive black holes (1663); Active galactic nuclei (16); High energy astrophysics (739) 
\end{abstract}


\section{Introduction} \label{sec:intro}

Active galactic nuclei (AGNs) are powered by accretion onto supermassive black holes (SMBHs).
A large fraction of the AGN radiation is in the optical/UV and \xray\ bands, likely originating from the accretion disk and \hbox{accretion-disk} corona, respectively.
A significant correlation between the AGN optical/UV emission and \xray\ emission has been observed across a broad range of AGN luminosity, typically expressed as a negative correlation between the \hbox{optical-to-X-ray} \hbox{power-law} slope parameter ($\alpha_{\rm OX}$)\footnote{$\alpha_{\rm OX}$ is defined as $\alpha_{\rm OX}=-0.3838~{\rm log}(f_{2500~\textup{\AA}}/f_{\rm 2~keV})$, where $f_{2500~\textup{\AA}}$ and $f_{\rm 2~keV}$ are the flux densities at \hbox{2500 \AA} and \hbox{2 keV}, respectively.} and the $2500~\textup{\AA}$ monochromatic luminosity ($L_{2500~\textup{\AA}}$; e.g., \citealt{Avni1982, Kriss1985, Avni1986, Wilkes1994, Vignali2003, Strateva2005, Steffen2006, Just2007, Gibson2008, Green2009, Grupe2010, Lusso2010, Jin2012, Chiaraluce2018, Laha2018, Liu2021}).
This correlation can also be described as a \hbox{non-linear} positive correlation between the \hbox{2 keV} luminosity ($L_{\rm 2~keV}$) and $L_{2500~\textup{\AA}}$ \citep[e.g.,][]{Tananbaum1979, Avni1982, Avni1986, Wilkes1994, Strateva2005, Steffen2006, Just2007, Lusso2010, Lusso2016, Chiaraluce2018}.
These relations indicate an intrinsic \hbox{disk-corona} connection, and they have been used to identify AGNs emitting unusually weak or strong \xray\ emission \citep[e.g.,][]{Gibson2008, Miller2011, Pu2020, Fu2022} or to estimate cosmological parameters \citep[e.g.,][]{Lusso2016, Bisogni2017, Lusso2017}.
 

The underlying physics for the empirical $\alpha_{\rm OX}\textrm{--}L_{2500~\textup{\AA}}$ relation remains elusive, and some studies have suggested that it may be driven by more fundamental correlations between $\alpha_{\rm OX}$ and the SMBH mass \hbox{and/or} Eddington ratio \citep[e.g.,][]{Steffen2006, Kubota2018, Arcodia2019, Cheng2020, Liu2021, Rankine2023}.
For example, \cite{Liu2021} proposed that $\alpha_{\rm OX}=-0.13{\rm log}(L_{\rm Bol}/L_{\rm Edd})-0.10{\rm log}M_{\rm BH}-0.69$, where $L_{\rm Bol}/L_{\rm Edd}$ and $M_{\rm BH}$ are the Eddington ratio and SMBH mass, respectively.
In addition, the observed $\alpha_{\rm OX}\textrm{--}L_{2500~\textup{\AA}}$ relation has significant dispersion ($\approx0.1\textrm{--}0.2$; e.g., \citealt{Steffen2006}).\footnote{The dispersion is typically quantified as the standard deviation of the $\Delta\alpha_{\rm OX}$ values in different luminosities bins, where $\Delta\alpha_{\rm OX}$ is the difference between the observed $\alpha_{\rm OX}$ and the expected one from the $\alpha_{\rm OX}\textrm{--}L_{2500~\textup{\AA}}$ relation: $\Delta\alpha_{\rm OX}=\alpha_{\rm OX}-\alpha_{\rm OX,~exp}$.}
The dispersion might have contributions from several possible effects: a) \hbox{X-ray} obscuration, b) UV extinction, c) intrinsic \hbox{X-ray} weakness, d) \hbox{host-galaxy} contamination in the UV, e) \hbox{non-simultaneous} \hbox{X-ray} and UV measurements.
Nevertheless, the majority of the dispersion appears intrinsic \citep[e.g.,][]{Vagnetti2010, Vagnetti2013}.
A few studies suggested that the correlation might have a luminosity dependence, as the slope of the relation appears steeper for luminous quasars \citep[e.g.,][]{Steffen2006, Timlin2020, Pu2020}.
There are also explorations of the \hbox{X-ray} and UV energy dependence (related to the $\alpha_{\rm OX}$ definition) of the relation \citep[e.g.,][]{Young2010, Jin2012, Signorini2023, Jin2024}.
The $\alpha_{\rm OX}$ parameter has at most a weak anticorrelation with redshift \citep[e.g.,][]{Steffen2006, Just2007, Green2009, Lusso2010, Vagnetti2010, Vagnetti2013, Chiaraluce2018}.

It is actually difficult to select a large sample of \hbox{type 1} AGNs across broad luminosity and redshift ranges to study the $\alpha_{\rm OX}\textrm{--}L_{2500~\textup{\AA}}$ relation and then explore the underlying \hbox{disk-corona} connection.
Optical/UV spectroscopic selection yields large samples of \hbox{type 1} AGNs, e.g., from the Sloan Digital Sky Survey (SDSS), but they usually lack good \hbox{X-ray} coverage and are limited to luminous quasars.
Deep \hbox{X-ray} surveys, such as the Chandra Deep Fields \citep[e.g.,][]{Brandt2005,Brandt2015,Xue2017}, provide a relatively complete AGN sample across the Universe, but the small areas limit the numbers of \hbox{type 1} AGNs available and it is also very expensive to obtain complete spectroscopic identifications.
There are also various systematics and biases to consider when combining samples from different surveys.
Therefore, the current studies of the $\alpha_{\rm OX}\textrm{--}L_{2500~\textup{\AA}}$ relation are either limited by the sample size (a few hundred AGNs) or the luminosity range (luminous quasars only),  hampering our ability to address the relevant questions above and understand the AGN \hbox{disk-corona} connection.

A promising approach to construct a large sample of AGNs for studying the \hbox{disk-corona} connection is to utilize \hbox{medium-depth} \hbox{large-area} \hbox{X-ray} surveys.
Currently, the \hbox{XMM-Spitzer} Extragalactic Representative Volume Survey (\hbox{XMM-SERVS}) provides a good balance between survey area and X-ray sensitivity \citep[e.g.,][]{Chen2018, Ni2021}.
The total area of the three survey fields, \hbox{XMM-LSS} ($\approx5.3~{\rm deg}^{2}$), \hbox{W-CDF-S} ($\approx4.6~{\rm deg}^{2}$),  and \hbox{ELAIS-S1} ($\approx3.2~{\rm deg}^{2}$), is $\approx13~{\rm deg}^{2}$.
The average \hbox{X-ray} flux limit reaches $9.8\times10^{-15}~{\rm erg}~{\rm cm}^{-2}~{\rm s}^{-1}$ in the \hbox{0.5--10 keV} band, which is $\approx0.5~{\rm dex}$ lower than that for the generally available \hbox{X-ray} coverage of  SDSS quasars \citep[e.g.,][]{Lusso2017, Lyke2020}.
For \hbox{type 1} AGN identification, although there are several spectroscopic surveys available with additional ones ongoing (e.g., see Section 5.1 of \citealt{Ni2021}), these are far from complete to provide a representative AGN sample.
On the other hand, the \hbox{XMM-SERVS} fields are covered by deep multiwavelength imaging surveys from radio to UV (e.g., Table 1 of \citealt{Chen2018}), and they also belong to the Deep Drilling Fields of the Legacy Survey of Space and Time (LSST) of the Vera C. Rubin Observatory that will have superb photometric monitoring for $\approx 10$ years \citep[e.g.,][]{Brandt2018, Scolnic2018}.
Thus it is feasible to perform photometric selection of quasars with the multiwavelength photometric data, aided with also morphological information obtained from the imaging data.

Recently, \hbox{machine-learning} techniques have been demonstrated powerful for selecting luminous quasars in \hbox{large-area} surveys.
Compared to traditional \hbox{template-fitting} methods, \hbox{machine-learning} approaches are computationally economical, and they are not sensitive to \hbox{cross-calibration} uncertainties among multiple survey data sets (e.g., \hbox{zero-point} offsets; \citealt{Salvato2019}).
Some popular \hbox{machine-learning} methods include traditional statistical techniques \citep[e.g.,][]{Richards2004,Bovy2011}, support vector machines \citep[e.g.,][]{Gao2008,Peng2012}, neural networks \citep[e.g.,][]{Yeche2010,Tuccillo2015,Chaini2023,Mart2023}, \hbox{Skew-QSO} method \citep{Yang2017,Yang2022}, and \hbox{XGBoost} (Extreme Gradient Boosting; \citealt{Chen2016}). 
Among these techniques, \hbox{XGBoost} has been widely used in classifications among quasars, galaxies, and stars \citep[e.g.,][]{Mirabal2016, Jin2019, Fu2021, Golob2021, Hughes2022, Calderone2024, Fu2024}.
The advantages of \hbox{XGBoost} include parallel processing, regularization that reduces model overfitting, and a \hbox{built-in} routine to handle missing values.
Typically, \hbox{machine-learning} identification of AGNs has been applied to select luminous quasars (e.g., $i<21.5$; e.g., \citealt{Jin2019, Fu2021}).
However, the method should also be applicable to lower luminosity quasars as the AGN infrared \hbox{(IR)--UV} spectral energy distributions (SEDs) do not have a strong dependence on luminosity \citep[e.g,][]{Richards2006, Krawczyk2013}.
Besides quasar selection, the \hbox{machine-learning} technique has also been applied to derive quasar photometric redshifts.
Compared to the traditional SED fitting approach, \hbox{machine-learning} algorithms are able to infer photometric redshifts with comparable quality \citep[e.g,][]{Sanchez2014, Beck2017, Tanaka2018, Schmidt2020}.
Owing to the featureless \hbox{power-law} shape of the quasar optical SED, it is important to include UV, \hbox{near-IR} (NIR), and IR data in the process.

In this study, we use the \hbox{XGBoost} algorithm to photometrically select quasars in the $\approx5.3~{\rm deg}^{2}$ \hbox{XMM-LSS} field.
We chose relatively conservative criteria to construct a reliable quasar sample, and then we derived photometric redshifts for objects without spectroscopic redshifts.
Finally, we selected a subsample of these quasars to study the $\alpha_{\rm OX}\textrm{--}L_{2500~\textup{\AA}}$ relation with the \hbox{XMM-SERVS} \hbox{X-ray} data.
This approach can be applied to the other two  \hbox{XMM-SERVS} fields to assemble a larger sample, which we defer to future work.
The paper is organized as follows.
In Section~\ref{sec:sample}, we describe the multiwavelength data and the construction of the parent, training, and  \hbox{blind-test} samples.
In Section~\ref{sec:selection}, we present our  \hbox{machine-learning} processes and the quasar selection results.
Photometric redshifts for our selected quasars are derived in Section~\ref{sec:photo_z}.
In Section~\ref{sec:disk_corona}, we study the \hbox{\aox--\ltkf} relation for a subsample of 1\,007 quasars.
We summarize our results in Section~\ref{sec:conclu_future}.
Throughout this paper, we adopt a cosmology of \hbox{$H_{0}=67.4~{\rm km}~{\rm s}^{-1}~{\rm {Mpc}}^{-1}$}, $\Omega_{\rm M} = 0.315$, and $\Omega_{\Lambda} = 0.685$ \citep{Planck2020}.
The 2500~\AA\ monochromatic luminosity (\ltkf) is expressed in units of ${\rm erg}~{\rm s}^{-1}~{\rm Hz}^{-1}$.

\section{Multiwavelength Photometry and Sample Construction} \label{sec:sample}

\subsection{Multiwavelength Photometric Catalogs} \label{subsec:photo_catalog}
The \hbox{XMM-LSS} field has superb multiwavelength photometric coverage spanning from radio to UV.
For our study, we primarily utilized the catalogs in the IR, NIR, optical, and UV bands.
All the magnitude measurements have been \hbox{PSF-corrected}.

The IR catalog utilized in this study is the Spitzer DeepDrill \hbox{XMM-LSS} \hbox{two-band} catalog \citep{Lacy_DeepDrill_twobands, Lacy2021}.\footnote{\url{https://irsa.ipac.caltech.edu/cgi-bin/Gator/nph-dd}.}
The Spitzer DeepDrill survey covers the entire \hbox{XMM-LSS} field, using the Spitzer Infrared Array Camera (IRAC) $3.6~\micron$ and $4.5~\micron$ filters.
This survey achieves a $5\sigma$ point-source depth of $2~\mu{\mathrm{Jy}}$ (corresponding to an AB magnitude of 23.1) in each of the two bands.
The full width at half maximum (FWHM) of the point spread function (PSF) for the IRAC $3.6~\micron$ band is approximately $2\arcsec$.
We adopted the Spitzer DeepDrill \hbox{1.9\arcsec-radius} aperture photometry at IRAC $3.6~\micron$ and $4.5~\micron$ bands in this study.

The NIR catalog is the VISTA Deep Extragalactic Observations survey (VIDEO; \citealt{Jarvis2013}) Data Release 5 (DR5) catalog.
Within the \hbox{XMM-Newton} field, the VIDEO survey covers an area of $\approx4.5~\mathrm{deg}^{2}$ in the $Z$, $Y$, $J$, $H$, and $K_{\mathrm{S}}$ bands.
This survey reaches a $5\sigma$ AB-magnitude depth of $K_{\mathrm{S}}=23.8$ with a \hbox{2\arcsec}-diameter aperture.
We adopted the VIDEO \hbox{2\arcsec-diameter} aperture photometry in the $J,~H,~{\rm and}~K_{S}$ bands in this study, and we only selected those objects with \hbox{{\sc sextractor}} flags $<=3$ \citep{Bertin1996}.

The optical catalog is the Hyper \hbox{Suprime-Cam} Subaru Strategic Programme Data Release 3 catalog ( \hbox{HSC-SSP} DR3; \citealt{Aihara2022}).\footnote{\url{https://hsc-release.mtk.nao.ac.jp/doc/index.php/faq__pdr3}.}
This catalog contains imaging data collected with five \hbox{broad-band} filters ($g$, $r$, $i$, $z$, and $y$).
Within the \hbox{XMM-LSS} field, \hbox{HSC-SSP} has three layers of surveys, including the Wide (covers the entire \hbox{XMM-LSS} field), Deep ($\approx5~{\rm deg}^{2}$), and UltraDeep ($\approx1.77~{\rm deg}^{2}$) surveys \citep{Aihara2018}.
The $5\sigma$ limiting magnitudes in the $i$ band for the HSC Wide survey and combined Deep/UltraDeep survey (processed jointly since DR2) are 26.2 and 26.9, respectively \citep{Aihara2022}.
The median FWHM of the \hbox{HSC-SSP} \hbox{$i$-band} PSF is $0{\arcsec}.69$ \citep{Miyazaki2018}.
In this study, we adopted the PSF magnitudes in the five bands to construct colors, and we used the differences between the PSF and Kron magnitudes (e.g., $i_{\rm PSF}-i_{\rm Kron}$) in the five bands as indicators of source morphology \citep[e.g.,][]{Jin2019,Fu2021}.

The UV catalogs are the \hbox{Canada-France-Hawaii} Telescope Legacy Survey (CFHTLS; \citealt{Hudelot2012}) and GALEX Deep Imaging Survey \citep{Martin2005} catalogs.
CFHTLS comprises two survey layers in the \hbox{XMM-LSS} field: a wide layer (CFHTLS-W1) covering the entire ${\rm 5.3~deg}^{2}$ \hbox{XMM-LSS} field, and a deep layer  (CFHTLS-D1) covering a 1 ${\rm deg}^{2}$ region.
The \hbox{CFHTLS-W1} catalog reaches an 80\% completeness limit of  $u*=25.2$ for point sources, and the \hbox{CFHTLS-D1} catalog has an 80\% completeness limit of  $u*=26.3$.
The average PSF FWHM for the CFHTLS $u*$ band is $\approx0{\arcsec}.85$.
The GALEX Deep Imaging Survey covers the entire \hbox{XMM-LSS} field.
Within this area, the 99\% limiting magnitude in the \hbox{near-UV} (NUV) band is $\approx25.3$. 
The 80\% PSF sizes in the GALEX NUV and \hbox{far-UV} (FUV) bands are both $\approx6{\arcsec}$.
We adopted the \hbox{3\arcsec-diameter} aperture photometry in the $u^{*}$ band from the CFHTLS catalogs and the calibrated fluxes in the FUV and NUV bands from the GALEX catalog.
We excluded CFHTLS and GALEX sources with \hbox{{\sc sextractor}} flags $>3$.
We also excluded those \galex\ sources labeled with ${\rm artifacts} =2~{\rm or}~4$ following the GALEX official instructions.\footnote{\url{https://galex.stsci.edu/GR6/?page=ddfaq\#6}.}

\subsection{Parent Sample Construction} \label{subsec:parent}
We selected our \hbox{parent-sample} sources from the HSC DR3 catalog.
We adopted additional filtering criteria based on data quality, IR detections, and source blending.

\begin{enumerate}

\item
We selected HSC sources from a circular region with a radius of 1\arcdeg.9 (an area of $\approx11~{\rm deg}^{2}$) centering at ${\rm RA}=35\arcdeg.5743$ and ${\rm DEC}=-4\arcdeg.7485$.
As shown in Figure~\ref{fig:fig1}, this region completely covers the \hbox{XMM-SERVS} \hbox{XMM-LSS} field.
This circular region is an approximation of the Spitzer DeepDrill survey field so that we have good  Spitzer IRAC photometric coverage.

\item
We set an \hbox{$i$-band} magnitude cut of $i<23$.
Given the 50\% \hbox{full-band} flux limit of $4.4\times10^{-15}~{\rm erg}~{\rm cm}^{-2}~{\rm s}^{-1}$ in the \hbox{XMM-SERVS} \hbox{XMM-LSS} survey \citep{Chen2018}, we estimate that quasars with $i<22.5$ and redshift up to $\approx2.3$ (corresponding to $\approx85\%$ of the \hbox{training-sample} quasars) can be detected, if they exhibit nominal levels of \xray\ emission as expected from the \hbox{\aox--\ltkf} relation of \cite{Just2007}.
We relaxed the criterion to $i<23$ in the quasar selection stage.

\item
Referring to the HSC official instructions, we selected HSC \hbox{Wide-layer} and \hbox{Deep/UltraDeep-layer} catalog sources with clean photometry by excluding those with bad pixels, cosmic rays in the source centers, pixels outside the usable exposure regions, or bad centroids.
We required each HSC source to have all the \hbox{5-band} photometric measurements in order to obtain its comprehensive colors.
To obtain reliable morphological features in each band (e.g., $i_{\rm PSF}-i_{\rm Kron}$) which is useful for distinguishing point sources (quasars and stars) from extended sources (galaxies), we also excluded sources with flags of problematic \hbox{PSF-magnitude} or \hbox{Kron-magnitude} measurements.
We merged the HSC \hbox{Wide-layer} and \hbox{Deep/UltraDeep-layer} catalogs using a \hbox{1\arcsec-matching} radius, with the \hbox{Deep/UltraDeep-layer} catalog having a higher priority.
We then obtained a sample of 259\,707 HSC sources with HSC $i<23$. 

\item
We required that every source must have Spitzer DeepDrill IRAC measurements, as \hbox{mid-infrared-related} colors are critical for quasar selection and photometric redshift estimation \citep[e.g.,][]{Jin2019, Fu2021}.
All type~1 quasars with {$i<23$} are expected to be detectable in the Spitzer DeepDrill survey considering the typical quasar SED in \cite{Krawczyk2013}.
We excluded Spitzer DeepDrill sources with flux densities smaller than their uncertainties or \hbox{{\sc sextractor}} flags $>3$ in either of the two IRAC bands.
We then matched the Spitzer DeepDrill sources to the HSC sources using a \hbox{1\arcsec-matching} radius and obtained 184\,140 matches.
There are 142 pairs where two HSC sources matched to one IRAC source.
For the rest 183\,856 (184\,140$-$284) matches, all HSC sources has a \hbox{one-to-one} match to a unique Spitzer DeepDrill source.
\item
The \hbox{HSC-SSP} survey has a better spatial resolution than the Spitzer DeepDrill survey (Section \ref{subsec:photo_catalog}), and thus one Spitzer DeepDrill IRAC source might be the blend of two neighbouring HSC sources.
We considered an HSC source as potentially IRAC blended if there are other $i < 25.5$ HSC sources within $2\arcsec$ and these $i < 25.5$ HSC sources have no other matched IRAC sources, and subsequently excluded it.
All of the 142 pairs with two HSC sources matched to one IRAC source are excluded in this filtering process.
This filtering process resulted in a sample of 168\,905 HSC sources, and all of them have a \hbox{one-to-one} match to an unique IRAC source.
\end{enumerate}

When matching the 168\,905 HSC sources to the VIDEO sources using a $1\arcsec$-matching radius, we found that there are 100 HSC sources with multiple VIDEO counterparts. 
We checked the HSC images of these sources and found that they are likely blended with nearby sources, which were not identified in the HSC source catalogs. 
We then excluded these sources and obtained 168\,805 sources.
These objects constitute our parent sample.

We show the spatial distribution of the \hbox{parent-sample} sources in Figure~\ref{fig:fig1}.
In Table~\ref{tbl:tbl1}, we list the fraction of our \hbox{parent-sample} sources with available HSC and Spitzer IRAC photometry.
For a small number ($\approx0.1\%$) of \hbox{parent-sample} sources, they do not have Kron photometry in one or more specific HSC bands and are still labeled as sources with Kron photometry in the HSC DR3 source catalog.
As a result, the fraction of \hbox{parent-sample} sources with available Kron photometry in a specific HSC band is not $100\%$.
The \hbox{Kron-magnitude-related} colors of sources without Kron photometry are treated as missing values in our selection stage (see Section~\ref{sec:selection} below).
The distribution of the \hbox{$i$-band} magnitudes of the \hbox{parent-sample} sources is shown in Figure~\ref{fig:imagdis}.
The \hbox{parent-sample} sources have an average \hbox{$i$-band} magnitude of 21.7.
All \hbox{parent-sample} sources are detected in the HSC $g,~r,~i,~z,~{\rm and}~y$ bands (with available PSF photometry) and Spitzer DeepDrill IRAC $3.6~\micron$ and $4.5~\micron$ bands, as per the sample construction.

\begin{deluxetable}{lcccc}
\tablewidth{0pt}
\tablecaption{Summary of Sample Photometry}
\tablehead{
\colhead{Survey}  &
\colhead{Parent}  &
\colhead{Training} &
\colhead{Blind-test} \\
\colhead{Band}  &
\colhead{Sample}  &
\colhead{Sample}  &
\colhead{Sample} \\
\colhead{(1)} &
\colhead{(2)} &
\colhead{(3)} &
\colhead{(4)}
}
\startdata
\galex\ FUV & 17.2\% & 18.3\% & 18.5\% \\
\galex\ NUV & 30.7\% & 49.6\% & 44.2\% \\
CFHTLS $u^{*}$ & 86.0\% & 99.3\% & 67.9\% \\
HSC $g$, $r$, $i$, $z$, and $y$ (PSF) &100.0\%& 100.0\%& 100.0\% \\
VIDEO $J$ & 50.8\% & 80.9\% & 16.8\% \\
VIDEO $H$ & 50.8\% & 80.9\% & 16.8\% \\
VIDEO $K_{S}$ & 50.8\% & 81.0\% & 16.8\%\\
DeepDrill IRAC [3.6] and [4.5] &100.0\%& 100.0\%& 100.0\% \\
HSC $g_{\rm Kron}$&99.9\%&99.9\%& 99.9\% \\
HSC $r_{\rm Kron}$&99.9\%&99.9\%& 99.9\% \\
HSC $i_{\rm Kron}$&99.9\%&99.9\%& 99.9\% \\
HSC $z_{\rm Kron}$&99.9\%&99.9\%& 99.9\% \\
HSC $y_{\rm Kron}$&99.8\%&99.9\%& 99.8\% \\
\enddata
\tablecomments{Column (1): name of the survey band. Columns (2)-(4): fraction of the parent sample, training sample, and blind-test sample sources with available photometry in each band, respectively. A small number ($\approx0.1\%$) of \hbox{parent-sample} sources do not have Kron photometry in one or more specific HSC bands and are still labeled as sources with Kron photometry in the HSC DR3 source catalog. Therefore, the fraction of \hbox{parent-sample} sources with available Kron photometry in a specific HSC band is not $100\%$.}
\label{tbl:tbl1}
\end{deluxetable}

\subsection{VIDEO NIR, CFHTLS $u$-band and GALEX UV Photometry} \label{subsec:uv_data}

\begin{figure}
\centering
\includegraphics[scale=0.4]{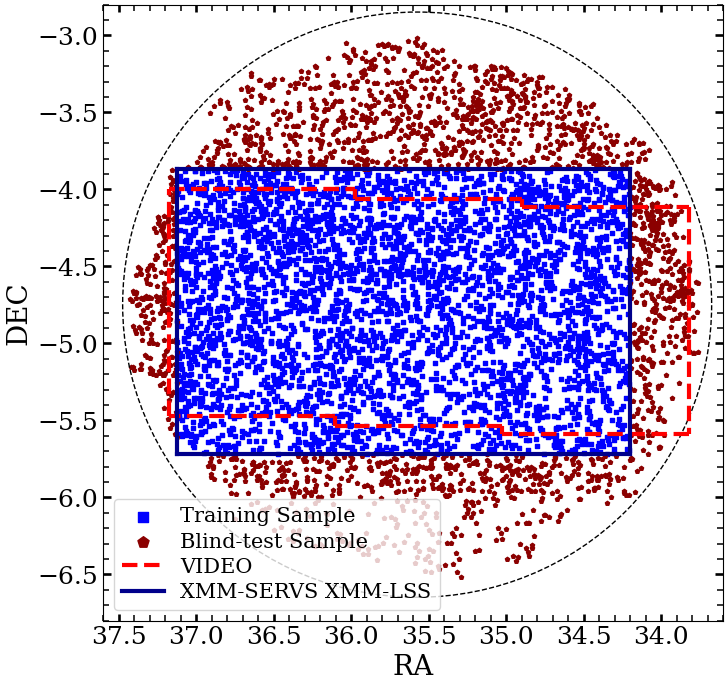}
\caption{
Spatial distributions of our sample sources.
The circle represents the region we used to select initial HSC catalog sources which is an approximation of the Spitzer DeepDrill field.
The red dashed line represents the VIDEO survey field.
The \hbox{dark-blue} solid line represents the \xray\ source catalog selection area in \cite{Chen2018}.
The blue squares and the \hbox{dark-red} pentagons represent the \hbox{training-sample} sources and \hbox{blind-test-sample} sources, respectively.
Individual parent sample sources are not shown, but except small cavities which may be caused by bright stars and bad pixels, the spatial distribution of our \hbox{parent-sample} sources is pretty uniform.
}
\label{fig:fig1}
\end{figure}

For each of the sources in our parent sample, the VIDEO $J$-, $H$-, $K_{S}$-band photometric measurements were collected by matching the VIDEO catalog sources to its HSC coordinate using a \hbox{1\arcsec-matching} radius.
In Table~\ref{tbl:tbl1}, we list the fraction of our \hbox{parent-sample} sources with available VIDEO photometry; only $\approx50\%$ of our \hbox{parent-sample} sources have available VIDEO photometry due to the limited coverage of the VIDEO survey field (see Figure~\ref{fig:fig1}).
All parent sample objects within the VIDEO footprint have VIDEO counterparts, except for a few sources that are located near bright stars or galaxies, which might have affected the VIDEO photometric extraction.

We searched for the $u^{*}$-band measurements from the CFHTLS source catalogs for the parent sample.
We matched the HSC \hbox{parent-sample} sources to the CFHTLS-D1 and CFHTLS-W1 sources using an \hbox{1\arcsec-matching} radius to obtain the \hbox{$u^{*}$-band} measurements, and the CFHTLS-D1 photometry is preferred.
In Table~\ref{tbl:tbl1}, we list the fraction of our \hbox{parent-sample} sources with available CFHTLS $u^{*}$-band photometry.

We collected the UV measurements from the \galex\ Deep Imaging Survey source catalog for our parent sample.
A source in the \hbox{XMM-LSS} field usually has multiple \galex\ measurements in the \galex\ source catalog.
Following a method similar to that adopted by \cite{Zou2022}, we selected unique measurements for each \galex\ source.
We first ranked all \galex\ catalog sources according to their detection status: sources with detection in both NUV and FUV bands were ranked at the top, while sources with only NUV or FUV detection were ranked in the middle and the bottom.
Then, for sources within each level of detection status, we assigned higher rankings to those with longer exposure times.
Among sources with the same detection status and exposure times, we assigned higher rankings to those with smaller $\mathtt{fov\_radius}$ values.
To construct our unique source catalog, we matched each GALEX source with a high rank to the GALEX sources with lower ranks with a matching radius of 2.5\arcsec, and we excluded the matched sources with lower ranks.
Then we matched the unique \galex\ sources to our \hbox{parent-sample} sources using an \hbox{1\arcsec-matching} radius.
In Table~\ref{tbl:tbl1}, we list the fraction of our \hbox{parent-sample} sources with available \galex\ photometry.


\subsection{Spectroscopic Training Sample}\label{subsec:spec_training}
From the parent sample, we selected spectroscopically confirmed quasars from the SDSS Data Release 16 quasar catalog \citep{Lyke2020} in the \hbox{XMM-LSS} field as the \hbox{training-sample} quasars.
We matched these SDSS DR16 quasars to the HSC coordinates of our \hbox{parent-sample} sources using a \hbox{1\arcsec-matching} radius and obtained 1\,260 matched quasars.
To ensure that the spectroscopically confirmed quasars are indeed \hbox{type 1} quasars with reliable redshift measurements, we selected 1116 SDSS quasars with no pipeline warning on its redshift measurement ($\mathtt{zwarning=0}$).

From sources in the SpecPhotoAll table of SDSS DR16, we utilized CASJOB to extract a sample of spectroscopically identified galaxies and stars with $\mathtt{zwarning=0}$.
Subsequently, we matched these galaxies and stars to our \hbox{parent-sample} sources using a \hbox{1\arcsec-matching} radius.
We then selected those \hbox{parent-sample} sources with matched SDSS spectra inside the \hbox{XMM-SERVS} \hbox{XMM-LSS} field as our training sample.
Our training sample comprises 723 quasars, 2\,673 galaxies, and 314 stars.
We list these numbers in Table~\ref{tbl:n_of_source1}.
The spatial distribution of the training sample objects is illustrated in Figure~\ref{fig:fig1}.

\begin{figure}
\centering
\includegraphics[scale=0.45]{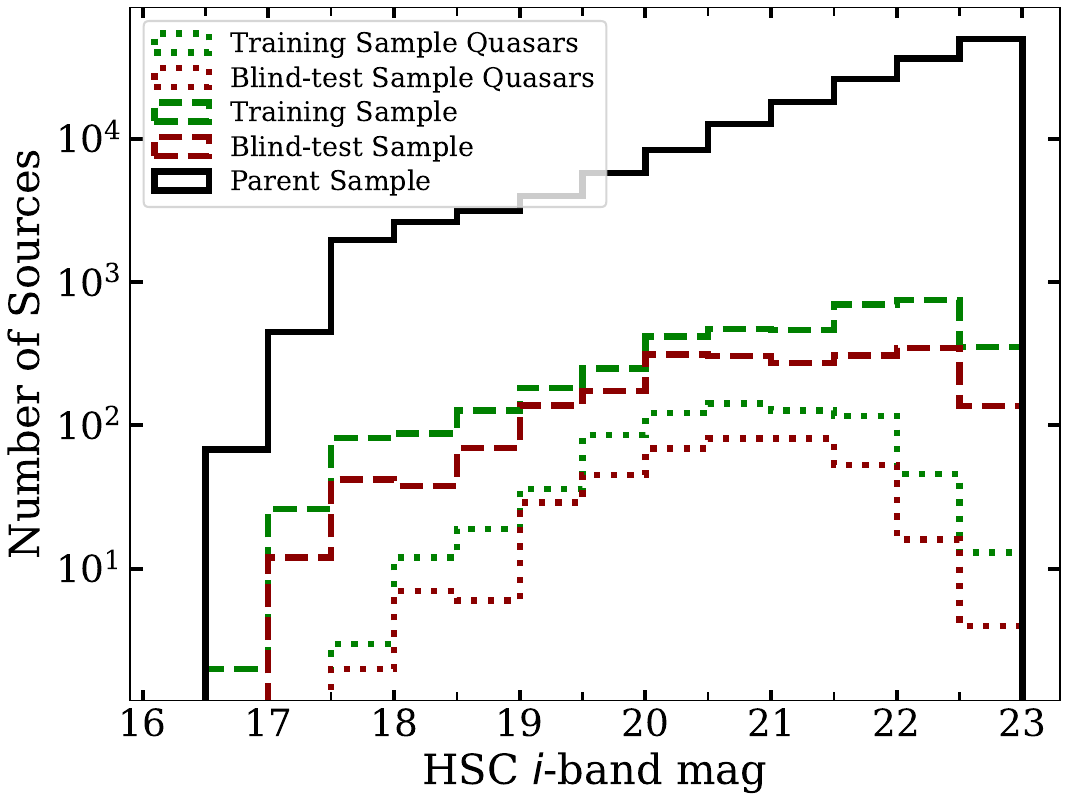}
\caption{
Distribution of the observed \hbox{$i$-band} magnitude of our \hbox{parent-sample}, \hbox{training-sample}, and \hbox{blind-test-sample} sources.
The training and \hbox{blind-test-sample} sources are brighter than the \hbox{parent-sample} sources, which is attributed to the target selection strategies utilized in SDSS spectroscopic observations.
}
\label{fig:imagdis}
\end{figure}

We display the distribution of the observed $i$-band magnitudes for the \hbox{training-sample} sources in Figure~\ref{fig:imagdis}.
With an average $i$-band magnitude of 20.7, the training sample quasars are brighter than the parent sample with an average $i$-band magnitude of 21.7.
This is because optically bright sources are preferred in SDSS spectroscopic observations.
Details regarding the fraction of our \hbox{training-sample} sources with available photometry are listed in Table~\ref{tbl:tbl1}.
Compared to the entire parent sample, our \hbox{training-sample} sources exhibit a significantly higher fraction of available photometry in the VIDEO $J,~H,$ and $K_{S}$ bands, primarily due to the coverage of the VIDEO survey field, which is in the \hbox{XMM-LSS} field (Figure~\ref{fig:fig1}).

\subsection{Blind-Test Sample} \label{subsec:spec_test}
In machine learning, models often capture patterns specific to the training sample, leading to high accuracy on the training sample but poor performance on new, unseen data. 
A \hbox{blind-test} sample, that is not included in the training procedure, is thus typically needed; it enables the evaluation of how well the model generalizes to unseen data.
The \hbox{blind-test} sample is essential for identifying issues such as overfitting, where a model performs well on the training data but poorly on new data.
Therefore, we created a \hbox{blind-test} sample from the \hbox{parent-sample} sources located outside the \hbox{XMM-SERVS} \hbox{XMM-LSS} field.

The \hbox{blind-test} sample was constructed from the SDSS DR16 catalogs in the same manner as the training sample (Section~\ref{subsec:spec_training}), resulting in 393 quasars, 1\,503 galaxies, and 155 stars.
We list these numbers in Table~\ref{tbl:n_of_source1}.
The spatial distribution of the \hbox{blind-test} sample objects is displayed in Figure~\ref{fig:fig1}.
We present the distribution of the observed \hbox{$i$-band} magnitudes for the \hbox{blind-test} sample sources in Figure~\ref{fig:imagdis}. 
The distribution of observed \hbox{$i$-band} magnitudes for our \hbox{blind-test-sample} quasars aligns with that of our training-sample quasars, both having an average \hbox{$i$-band} magnitude of 20.7.
We present the fraction of available photometry in each band for our \hbox{blind-test-sample} sources in Table~\ref{tbl:tbl1}.
When compared to our \hbox{training-sample} sources, the blind-test-sample sources show significantly lower fractions of available CFHTLS $u^{*}$ and VIDEO $J$, $H$, $K_{S}$ photometry.
All \hbox{blind-test-sample} sources are located outside the \hbox{XMM-LSS} field, and only a few of them have VIDEO coverage.
Furthermore, \hbox{blind-test-sample} sources with declinations larger than $-3\arcdeg.705$ also lack CFHTLS coverage.


\section{Photometric Selection of Quasars} \label{sec:selection}
In this study, we utilized \hbox{XGBoost} to construct our quasar selection procedure.
We developed two classifiers: the first to differentiate extragalactic objects (quasars and galaxies) from stars, and the second to specifically identify quasars while excluding galaxies.
In the first classifier, quasars and galaxies are grouped together based on the rationale that quasars are essentially a unique subclass of galaxies.
In principle, a single \hbox{three-class} classifier is capable of performing similar selections.
However, the \hbox{two-classifier} approach could reduce the impact of class imbalance (see discussion in Section 3.3 of \citealt{Fu2021}), making it easier to \hbox{fine-tune} the hyperparameters for underrepresented classes (i.e., quasars in the case here).
It also allows each binary classifier to use tailored features or thresholds that might not apply universally in a \hbox{three-class} classifier, which may improve model accuracy for each step.
The sample used to train the first classifier includes 314 stars and 3\,396 extragalactic objects; for the second classifier, the training sample consists of the 3\,396 extragalactic objects with 723 quasars and 2\,673 galaxies (see Section~\ref{subsec:spec_training}).

\subsection{Input Features} \label{subsec:features}
Our sample sources were characterized by input features consisting of 18 colors (${\rm FUV}-{\rm NUV}$, ${\rm NUV}-u^{*}$, ${\rm NUV}-g$, $u^{*}-g$, $g-r$, $r-i$, $i-z$, $z-y$, $y-J$, $J-H$, $H-K_{S}$, $K_{S}-{\rm IRAC}~[3.6]$, ${\rm IRAC}~~[3.6]-[4.5]$, $g-{\rm IRAC}~[3.6]$, $r-{\rm IRAC}~[3.6]$, $i-{\rm IRAC}~[3.6]$, $z-{\rm IRAC}~[3.6]$, and $y-{\rm IRAC}~[3.6]$) and 5 morphological parameters ($g_{\rm PSF}-g_{\rm Kron}$, $r_{\rm PSF}-r_{\rm Kron}$, $i_{\rm PSF}-i_{\rm Kron}$, $z_{\rm PSF}-z_{\rm Kron}$, and $y_{\rm PSF}-y_{\rm Kron}$).
Besides the first 13 independent colors derived from the photometry of neighboring bands (i.e., ${\rm FUV}-{\rm NUV}$ to ${\rm IRAC}~~[3.6]-[4.5]$),\footnote{The color ${\rm NUV}-g$ is used because not all sources have \hbox{$u^{*}$-band} photometry.} we also included the latter five colors to approximate the \hbox{optical-to-mid-IR (MIR)} SED shape that appeared useful for distinguishing quasars from stars in previous studies \citep[e.g.,][]{Jin2019,Fu2021}.
In cases where a source lacks available photometry or has a magnitude error greater than 0.2 in a given band, we denoted the magnitude in that band and its related colors with missing values (represented as ``999'').
\hbox{XGBoost} incorporates a \hbox{built-in} routine to handle these missing values (see Section~3.4 of \citealt{Chen2016}).
During tree construction, \hbox{XGBoost} assigns a default direction for missing values, allowing it to decide whether to send missing data to the left or right child node based on the optimal gain from that split.
This process can be interpreted as XGBoost learning an optimal imputation value
for missing data.\footnote{\url{https://github.com/dmlc/xgboost/issues/21.}}
To correct for Galactic extinction, we utilized the Milky Way \hbox{$R_{V}$-dependent} extinction model by \cite{Fitzpatrick2019} with $R_{V}=3.1$, along with Galactic extinction $E(B-V)$ values obtained from the dust map by \cite{Schlegel1998}.

\subsection{Model Optimization} \label{subsec:modelopt}
We utilized \textsc{optuna} \citep{Akiba2019}, an \hbox{open-source} hyperparameter optimization framework, to \hbox{fine-tune} the hyperparameters of \hbox{XGBoost} and obtain the optimal model.
The fixed hyperparameters of the \hbox{XGBoost} classifiers are as follows: $\mathtt{colsample\_bytree=1}$, $\mathtt{n\_estimators=100}$, $\mathtt{objective}=\mathtt{binary:logistic}$,
$\mathtt{seed}=0$,
$\mathtt{subsample=1}$, and $\mathtt{tree\_method}=\mathtt{hist}$.
These hyperparameters are set to either their default values or values referenced from previous studies \citep[e.g.,][]{Jin2019, Fu2021}.
Other hyperparameters were allowed to vary within a grid of values.
For example, the hyperparameter $\mathtt{max\_depth}$,\footnote{\url{https://xgboost.readthedocs.io/en/stable/parameter.html.}} which defines the maximum depth of a tree, is allowed to vary between 5 and 10. This range promotes simpler models that are less likely to overfit.

We utilized a \hbox{five-fold} \hbox{cross-validation} strategy to find the optimal hyperparameter sets for the two classifiers.
The entire training sample is used for both training and validation.
During the \hbox{five-fold} cross validation, the random seed is arbitrarily fixed at 100 for reproducibility.
We \hbox{fine-tuned} the model hyperparameters through 5\,000 trials to identify the optimal hyperparameter set for each classifier that minimized the average log loss value.
The log loss value is defined as:
\begin{equation}
	L_{\mathrm{log}}(y,p) = -{\frac{1}{N}}{\sum_{i=1}^{N}}(y\log(p) + (1-y)\log(1-p)).
\end{equation}
Here, $N$ is the number of sample sources, $y\in\{0,1\}$ represents the true label of a sample, and $p=\mathrm{Pr}(y=1)$ is the XGBoost-predicted probability estimate of a sample belonging to extragalactic objects or quasars.
The log loss value measures the accuracy of the predicted probabilities compared to the actual class labels (0 or 1) for binary classification.
The smaller the log loss value is, the better the model performs.
For the first and second classifiers, the average log loss values for the optimal hyperparameter sets during the \hbox{five-fold} cross validation are 0.010 and 0.054, respectively.

We also used two metrics, reliability and completeness, to evaluate the performance of the classifiers, which are defined as follows:
	\begin{equation}
		{\rm Reliability}=\frac{{\rm TP}}{{\rm TP}+{\rm FP}},
	\end{equation}
	\begin{equation}
		{\rm Completeness}=\frac{{\rm TP}}{{\rm TP}+{\rm FN}},
	\end{equation}
where true positive is denoted as TP, true negative as TN, false positive as FP, and false negative as FN.
In the first and second classifiers, TP refers to the correctly predicted extragalactic objects and quasars, respectively.
The reliability and completeness values during the \hbox{five-fold} cross validation and the \hbox{blind-test} validation of each classifier are presented in Sections~\ref{subsec:ext_select} and \ref{subsec:quasar_select} below.

\begin{deluxetable*}{ccccccc}[htb!]
	\tabletypesize{\scriptsize}
	\tablewidth{0pt}
	\tablecaption{Numbers of Sources in Each Sample}
	\tablehead{
		\colhead{Sample} &
		\colhead{Sources} &
		\colhead{SDSS Quasars} &
		\colhead{SDSS Galaxies} &
		\colhead{SDSS Stars} &
		\colhead{Quasars Selected  by \hbox{XGBoost}}  \\
		\colhead{(1)}  &
		\colhead{(2)}  &
		\colhead{(3)}  &
		\colhead{(4)}  &
		\colhead{(5)}  &
		\colhead{(6)}  
	}
	\startdata
	Parent Sample& 168805 & 1116 & 4176 & 469 & 2784 \\ 
	&  &  &  &  & (including 992 SDSS quasars) \\ 
	Training Sample& 3710 & 723 & 2673 & 314 & 651 \\
	(in the XMM-LSS field) &  &  &  &  & (including 651 SDSS quasars) \\ 
	Blind-test Sample& 2051 & 393 & 1503 & 155 & 344 \\
	(outside the XMM-LSS field) &  &  &  &  & (including 343 SDSS quasars) \\ 
	\enddata
	\tablecomments{
		Column (1): the name of samples.
		Column (2): the number of sample sources.
		Columns (3)-(5): the number of SDSS quasars, galaxies, and stars.
		Column (6): the number of sources selected as quasars by \hbox{XGBoost}.
		Some SDSS quasars are not recovered by \hbox{XGBoost}, and one galaxy in the \hbox{blind-test} sample was misidentified as a quasar by \hbox{XGBoost}.
	}
	\label{tbl:n_of_source1}
\end{deluxetable*}

\begin{figure}
	\centering
	\includegraphics[scale=0.45]{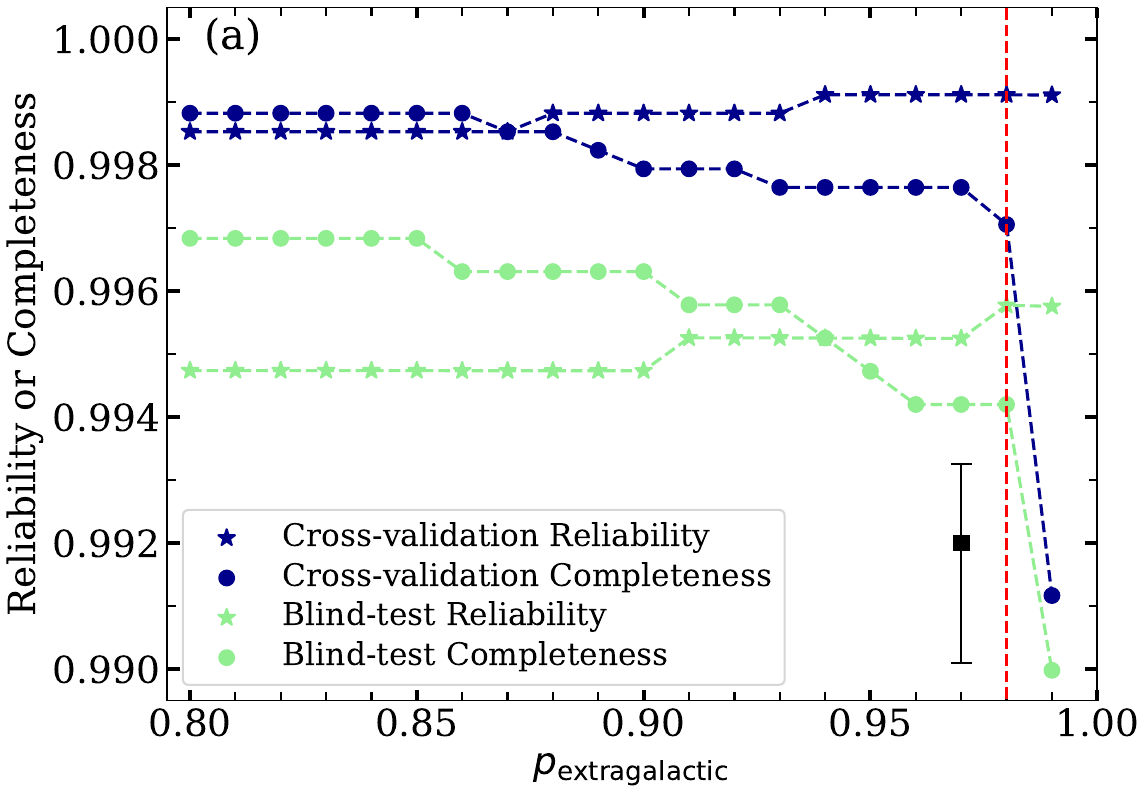}
	\includegraphics[scale=0.46]{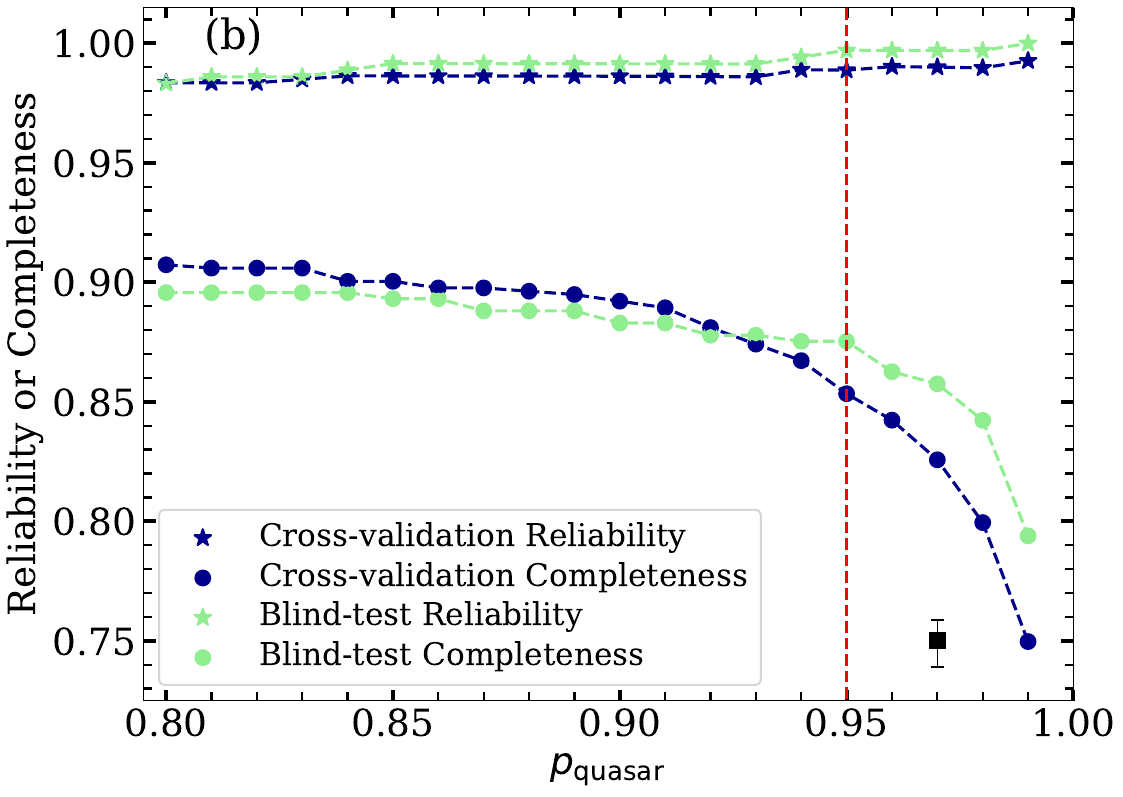}
	\caption{
		(a) Reliability and completeness of predicted  extragalactic objects for the \hbox{five-fold} \hbox{cross-validation} (dark blue points) and \hbox{blind-test} (light green points) results, plotted as a function of $p_{\rm extragalactic}$.
			(b)
			The same as panel (a), but for the predicted quasars from the second classifier.
			The $p_{\rm extragalactic}$ and $p_{\rm quasar}$ values range from 0.80 to 0.99, with increments of 0.01.
			The vertical red dashed lines indicate the thresholds used for classifying extragalactic objects ($p_{\rm extragalactic}>0.98$) and quasars ($p_{\rm quasar}>0.95$).
			The error bar in the lower right corner in each panel represents the median $1\sigma$ statistical uncertainty of the data points.
			The apparent large difference between the \hbox{cross-validation} and \hbox{blind-test} results in panel (a) is due to the small range of the \hbox{$y$-axis} values.
	}
	\label{fig:relia_comp_p}
\end{figure}

\begin{figure}
	\centering
	\includegraphics[scale=0.45]{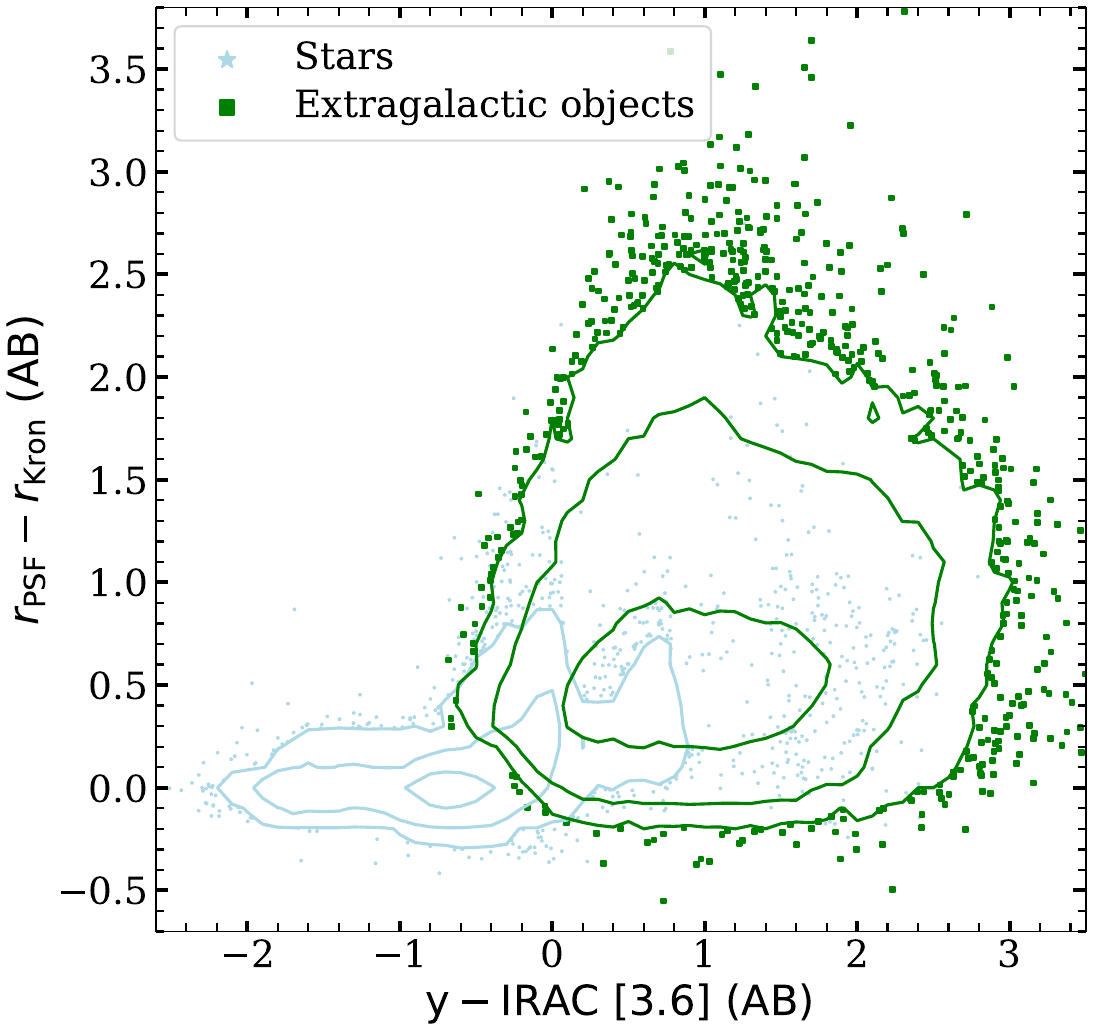}
	\caption{
		The color-morphology diagram of the stars (\hbox{light-blue} contours) and extragalactic objects (green contours) selected from our parent sample.
		Contours represent source number densities per bin, with a bin size of $0.1\times0.1$ along each axis.
		For extragalactic objects, the contour levels from the outside to the inside are spaced as 5, 50, and 500 per bin, while for stars, the contour levels are spaced as 10, 100, and 1000 per bin.
	}
	\label{fig:clf1}
\end{figure}

\begin{figure}
	\centering
	\includegraphics[scale=0.45]{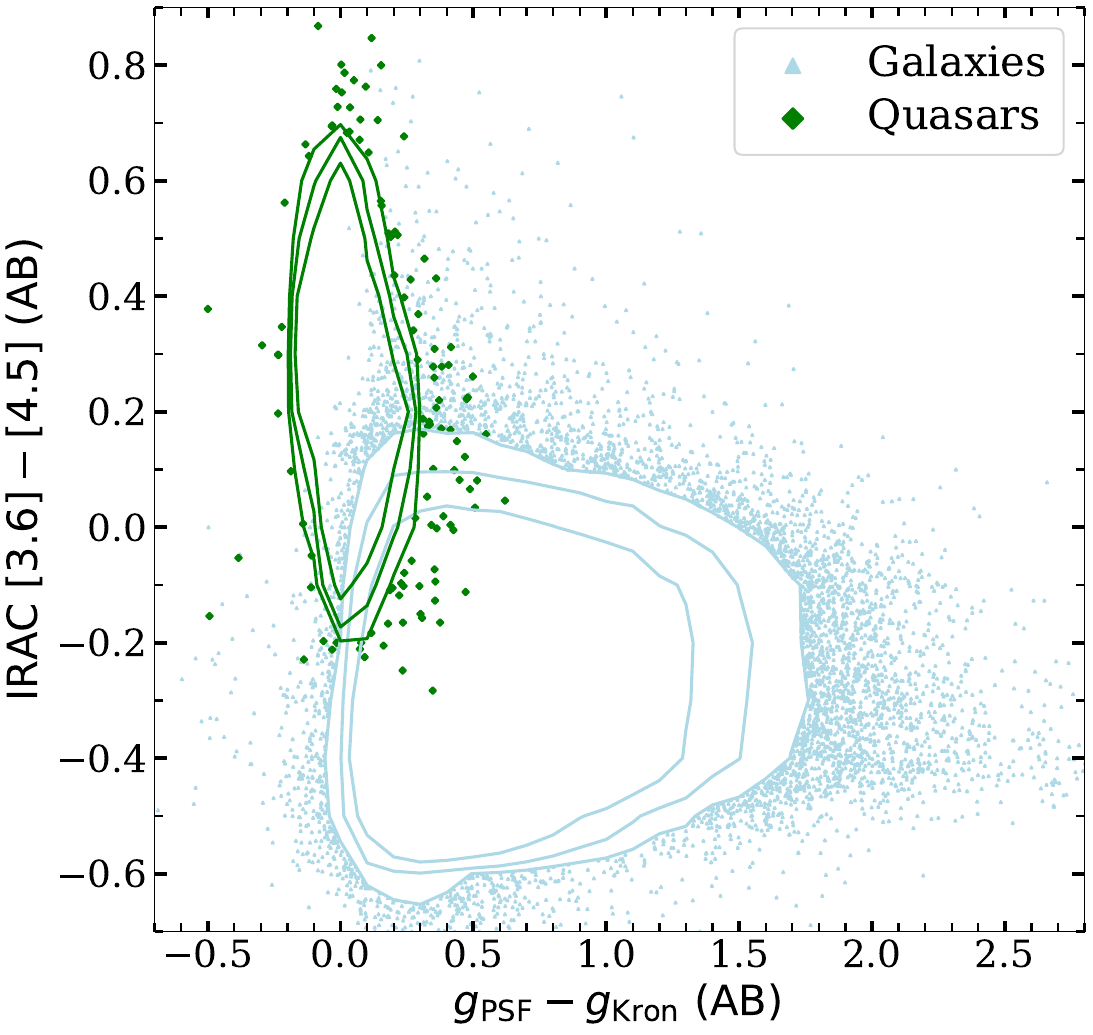}
	\caption{
		The morphology-color diagram of the galaxies (\hbox{light-blue} contours) and quasars (green contours) selected from our parent sample.
		Contours represent source number densities per bin, with a bin size of $0.1\times0.1$ along each axis.
		For quasars, the contour levels from the outside to the inside are spaced as 8, 16, and 32 per bin, while for galaxies, the contour levels are spaced as 150, 300, and 600 per bin.
	}
	\label{fig:clf2}
\end{figure}

\begin{deluxetable}{lcc}
\tablewidth{0pt}
\tablecaption{Performance of Our Selection
	\label{tbl:tbl3}}
\tablehead{
\colhead{Classes}  &
\colhead{Reliability}  &
\colhead{Completeness} \\
\colhead{(1)}  &
\colhead{(2)}  &
\colhead{(3)}
}
\startdata
Quasars & $0.997^{+0.002}_{-0.007}$ & $0.875^{+0.017}_{-0.019}$  \\
Galaxies & $0.966^{+0.004}_{-0.006}$ & $0.995^{+0.002}_{-0.003}$ \\
Stars & $0.930^{+0.021}_{-0.026}$ & $0.948^{+0.018}_{-0.024}$ \\
\enddata
\tablecomments{
The reliability and completeness of the \hbox{XGBoost} selected quasars, galaxies, and stars in the \hbox{blind-test} sample, calculated from the classification results of \hbox{XGBoost}.
The errors are $1\sigma$ statistical uncertainties calculated following Equations 16 and 17 of \citep{Gehrels1986}.}
\end{deluxetable}

\subsection{Selection of Extragalactic Objects} \label{subsec:ext_select}

To calculate the reliability and completeness for the first classifier, we need to set a threshold on the probability ($p_{\rm extragalactic}$) provided by \hbox{XGBoost} to determine the numbers of TP, TN, FP, and FN.
Therefore, we investigated the relations between the two metrics (reliability and completeness) of predicted extragalactic objects and $p_{\rm extragalactic}$ from the \hbox{five-fold} \hbox{cross-validation} results, as shown in Figure~\ref{fig:relia_comp_p}(a).
Additionally, we applied the first classifier to the \hbox{blind-test-sample} objects to examine these relations, also shown in Figure~\ref{fig:relia_comp_p}(a).
The representative $1\sigma$ statistical uncertainty is shown in the plot, derived following Equations 16 and 17 of \cite{Gehrels1986}.
Since the \hbox{blind-test-sample} objects were not used to train the classifier, their relations provide a suitable means for evaluating the true performance of the classifier on unseen data.
The \hbox{cross-validation} and \hbox{blind-test} results agree well (note the small \hbox{$y$-axis} range in the plot), and as expected, when $p_{\rm extragalactic}$ increases, the reliability increases but the completeness decreases.
Our primary objective in this study is to reliably select a sample of \hbox{type 1} quasars based on their photometry and investigate their \hbox{disk-corona} connection, and thus we need to minimize the biases introduced by stars and galaxies misclassified as quasars.
To exclude as many stars as possible and achieve a high completeness of extragalactic objects, we set $p_{\rm extragalactic}\geq0.98$ as the threshold to classify extragalactic objects.
Beyond this value, the reliability does not improve, but the completeness drops significantly.
During the \hbox{five-fold} cross validation, the reliability and completeness of extragalactic objects are 0.9991 and 0.9997, respectively.
For stars, the reliability and completeness are 0.9688 and 0.9904, respectively.
For the \hbox{blind-test} sample, the reliability and completeness of extragalactic objects are 0.9958 and 0.9942, respectively, while for stars, they are 0.9304 and 0.9484, respectively.
The \hbox{blind-test} results for stars are also listed in Table~\ref{tbl:tbl3}.

We then applied the first classifier to all \hbox{parent-sample} objects.
135\,272 of the \hbox{parent-sample} sources were classified as extragalactic objects, including all of the 723 \hbox{training-sample} quasars quasars.
Among the input features, the color $y-{\rm IRAC}~[3.6]$ and the morphology feature $r_{\rm PSF}-r_{\rm Kron}$ carried the highest and \hbox{second-highest} weights, respectively.
In Figure~\ref{fig:clf1}, we display our selected extragalactic objects and stars in the $y-{\rm IRAC}~[3.6]$ versus $r_{\rm PSF}-r_{\rm Kron}$ \hbox{color-morphology} feature space.
Our selected extragalactic objects generally exhibit redder $y-{\rm IRAC}~[3.6]$ colors compared to the selected stars.
Additionally, the selected extragalactic objects tend to have larger $r_{\rm PSF}-r_{\rm Kron}$ values on average, primarily because most of them are galaxies (see Section~\ref{subsec:quasar_select} below).
These features highlight the importance of MIR photometry and morphology in distinguishing extragalactic objects from stars \citep[e.g.,][]{Jin2019, Fu2021, Fu2024}.


\subsection{Selection of Quasars} \label{subsec:quasar_select}

For the second classifier, we also investigated the relation between the two metrics (reliability and completeness) of predicted quasars and $p_{\rm quasar}$, as shown in Figure~\ref{fig:relia_comp_p}(b).
We also applied the second classifier to the quasars and galaxies in the \hbox{blind-test} sample to examine these relations, as shown in Figure~\ref{fig:relia_comp_p}(b).
The two sets of metric values are overall consistent with each other.
To create a reliable sample of quasars and strike a balance between reliability and completeness, we adopted $p_{\rm quasar} \geq 0.95$ as the threshold to classify quasars, referring to the relations of the \hbox{blind-test} sample.
During the \hbox{five-fold} cross validation, the reliability and completeness of quasars are 0.9888 and 0.8534, respectively.
For galaxies, the reliability and completeness are 0.9618 and 0.9974, respectively.
For the \hbox{blind-test} sample, the reliability and completeness of quasars are 0.9971 and 0.8753, respectively, while for galaxies, they are 0.9658 and 0.9947, respectively.
The \hbox{blind-test} results for galaxies and quasars are also listed in Table~\ref{tbl:tbl3}.

We then applied the second classifier to the 135\,272 extragalactic objects classified by the first classifier, and selected 2\,784 quasars.
We list the number of quasars selected by \hbox{XGBoost} in Table~\ref{tbl:n_of_source1}.
We selected 1\,591 quasars within the \hbox{XMM-SERVS} \hbox{XMM-LSS} field (Figure~\ref{fig:fig1}) from the 2\,784 quasars to study their \hbox{disk-corona} connection (see Section~\ref{sec:disk_corona} below), given their \hbox{\xray} coverage in this field.
These 1\,591 quasars include 651 of the SDSS \hbox{training-sample} quasars.
The other 72 \hbox{training-sample} quasars were not classified as quasars and they were not included in our study of the \hbox{disk-corona} connection in Section~\ref{sec:disk_corona} below.
These 72 quasars were not correctly classified in  this step mainly because we adopted a strict threshold cut (i.e., $p_{\rm quasar}\geq 0.95$) to ensure high reliability of the selected quasars; many of these 72 quasars have relatively extended morphology, suggestive of higher \hbox{host-galaxy} contamination.
Using a lower threshold cut, such as $p_{\rm quasar}>0.2$, would include all these 72 quasars.
However, this would also increase significantly the numbers of quasars with strong \hbox{host-galaxy} contamination and galaxies misclassified as quasars, potentially biasing the derived \hbox{\aox--\ltkf} relation.

In the second classifier model, the morphology feature $g_{\rm PSF}-g_{\rm Kron}$ and the color ${\rm IRAC}~[3.6]-[4.5]$ carried the highest and \hbox{second-highest} weights.
In Figure~\ref{fig:clf2}, we illustrate the distribution of the selected quasars and galaxies in the $g_{\rm PSF}-g_{\rm Kron}$ versus ${\rm IRAC}~[3.6]-[4.5]$ \hbox{morphology-color} feature space.
Compared to our selected galaxies, our chosen quasars are more likely to be \hbox{point-like} sources, and they exhibit redder ${\rm IRAC}~[3.6]-[4.5]$ colors which are a characteristic feature of typical AGN SEDs.
These results highlight the importance of MIR colors and morphology in distinguishing quasars from galaxies in the photometric classification approach \citep[e.g.,][]{Fu2021, Golob2021, Fu2024}, offering valuable insights for future \hbox{multi-wavelength} surveys.



\subsection{Selection Reliability and Completeness} \label{subsec:relia_compl}


The \hbox{blind-test-sample} objects (see Section~\ref{subsec:spec_test}) were not included in the training of the quasar selection model, allowing us to use them to evaluate the performance of our selection.
The reliability and completeness values for the three types of objects from the \hbox{blind-test} results are summarized in Table~\ref{tbl:tbl3}.
The results are presented in Table~\ref{tbl:tbl3}.
The quasar selection is highly reliable; of the 345 selected quasars, 344 are SDSS quasars (Table~\ref{tbl:n_of_source1}) and only one galaxy was misidentified as a quasar, yielding a reliability of 0.997 (344/345).
The quasar selection completeness is 0.875 (344/393), still acceptable.
The completeness for the selected quasars is lower than that for the selected galaxies because we utilized a strict threshold for $p_{\rm quasar}$ in this study.
Some quasars, particularly those that are outshined by their host galaxies, may be classified as galaxies.
Utilizing a lower threshold would increase the completeness but 
reduce the reliability, resulting in more galaxies 
misclassified as quasars that may hamper the study of the \hbox{\aox--\ltkf} relation (galaxies are generally not luminous \xray\ emitters).
Our primary objective in this study is to reliably select a sample of \hbox{type 1} quasars based on their photometry and investigate their \hbox{disk-corona} connection, making high reliability a top priority. 
The VIDEO $J, H, K_{S}$ and CFHTLS $u^{*}$ coverage for the \hbox{blind-test} sample is less extensive than that for the parent sample in the \hbox{XMM-LSS} field (Table~\ref{tbl:tbl1}), so it is possible that the selection reliability and completeness for the \hbox{parent-sample} sources in the \hbox{XMM-LSS} field might be better.

\begin{figure}
\centering
\includegraphics[scale=0.45]{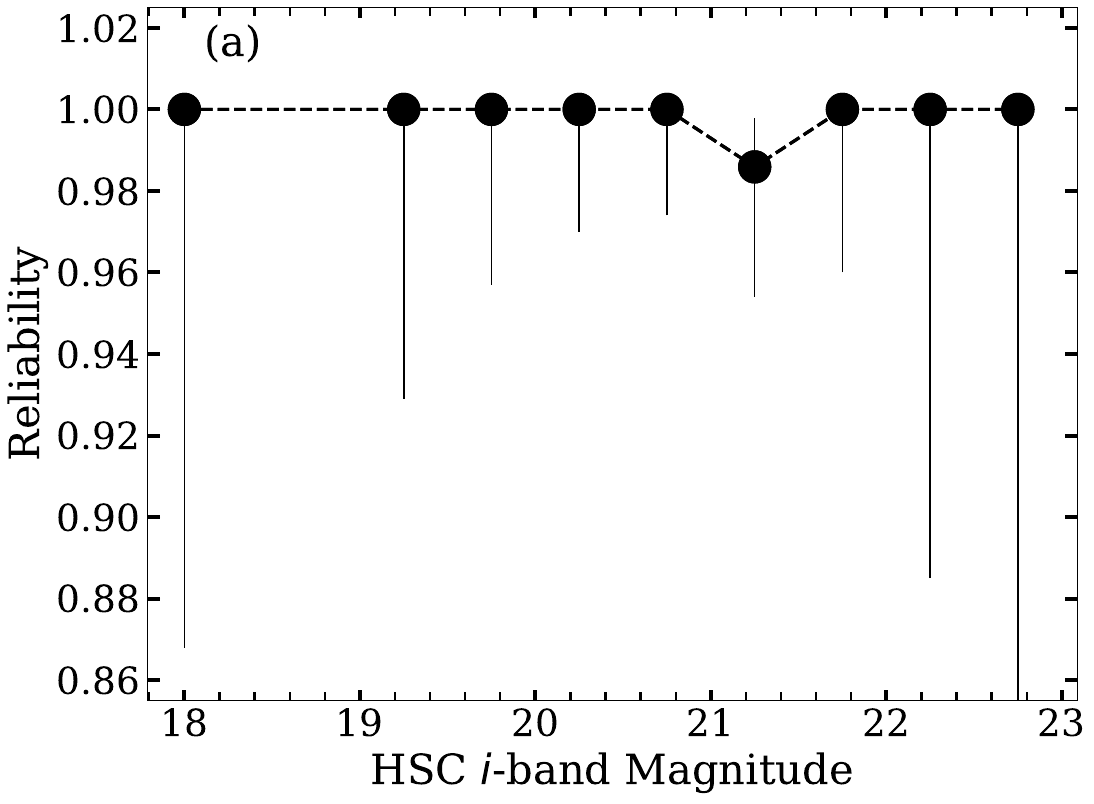}
\includegraphics[scale=0.45]{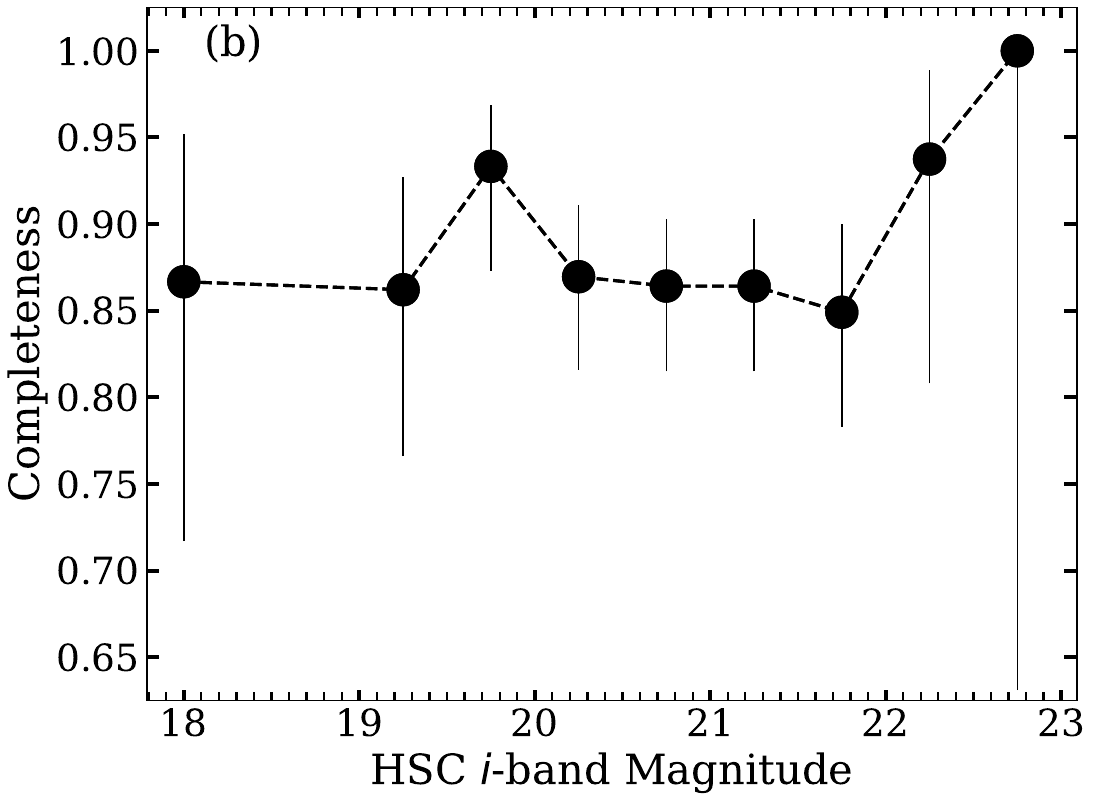}
\caption{
(a) Reliability versus observed \hbox{$i$-band} magnitude of the \hbox{blind-test-sample} quasars selected by \hbox{XGBoost}.
(b) Completeness versus observed \hbox{$i$-band} magnitude of the blind-test sample quasars.
The data are binned with a step of 0.5 magnitudes, except for those with $i<19$, which are grouped into one bin.
The errors are $1\sigma$ statistical uncertainties for individual bins \citep{Gehrels1986}.
There are no significant variations when considering the uncertainties.
}
\label{fig:fig5}
\end{figure}

One of the objectives of our study is to efficiently select a large sample of quasars that extend to lower luminosities compared to previous \hbox{machine-learning} results (see Section~\ref{sec:intro}).
Therefore, we examined the evolution of reliability and completeness as a function of the observed \hbox{$i-$band} magnitude of our \hbox{blind-test-sample} quasars selected by \hbox{XGBoost}, as shown in Figure~\ref{fig:fig5}.
No significant evolution was observed in the reliability or completeness parameters, probably because the \hbox{IR-to-UV} SED shapes of quasars do not have a strong dependence on luminosity \citep[e.g,][]{Richards2006, Krawczyk2013}.
Compared to previous studies that primarily selected optically brighter quasars for training and identification (e.g., \citealt{Jin2019} with $r\lesssim21.5$ and \citealt{Fu2021} with $i\lesssim22$), our approach is capable of identifying fainter quasars down to $i\approx22.5$; within the magnitude range of $22<i<23$, we selected 649 quasars ($\approx23\%$ of the 2\,784 selected quasars).

\subsection{Sky Density of Selected Quasars} \label{subsec:select_result}

The sky density of the 2\,784 \hbox{XGBoost} selected quasars is $253~{\rm deg}^{-2}$ ($2784/11~{\rm deg}^{-2}$), which is higher than that of the SDSS DR16 quasars in the same region ($101~{\rm deg}^{-2}$; $1116/11~{\rm deg}^{-2}$).
Our quasar sky density is comparable to the sky densities of quasars obtained from spectroscopic identifications of \xray\ AGNs in surveys with similar depths; for example, the broad-line AGN density is $\approx220~{\rm deg}^{-2}$ in the \hbox{XMM-Newton} COSMOS field \citep[e.g.,][]{Brusa2010}, and it is $\approx290~{\rm deg}^{-2}$ in the Chandra COSMOS field \citep[e.g.,][]{Marchesi2016}.

For the 1\,591 selected quasars within the \hbox{XMM-SERVS} \hbox{XMM-LSS} field (Figure~\ref{fig:fig1}) that are used to study the \hbox{disk-corona} connection (see Section~\ref{sec:disk_corona} below), the sky density reaches $300~{\rm deg}^{-2}$ ($1591/5.3~{\rm deg}^{-2}$), higher than that ($253~{\rm deg}^{-2}$) in the parent sample.
The higher density is likely due to the more extensive coverage of the VIDEO and CFHTLS surveys in this field.

\subsection{Spectroscopic Validation}
\label{subsec:spec_validation}
Among the 1\,591 selected quasars in the \hbox{XMM-SERVS} \hbox{XMM-LSS} field, we have obtained spectroscopic observations of 96 quasars using optical telescopes, including the MMT (Hectospec) \hbox{6.5-m} telescope and the WIYN (Hydra) \hbox{3.5-m} telescope.

\begin{figure}
	\centering
	\includegraphics[scale=0.25]{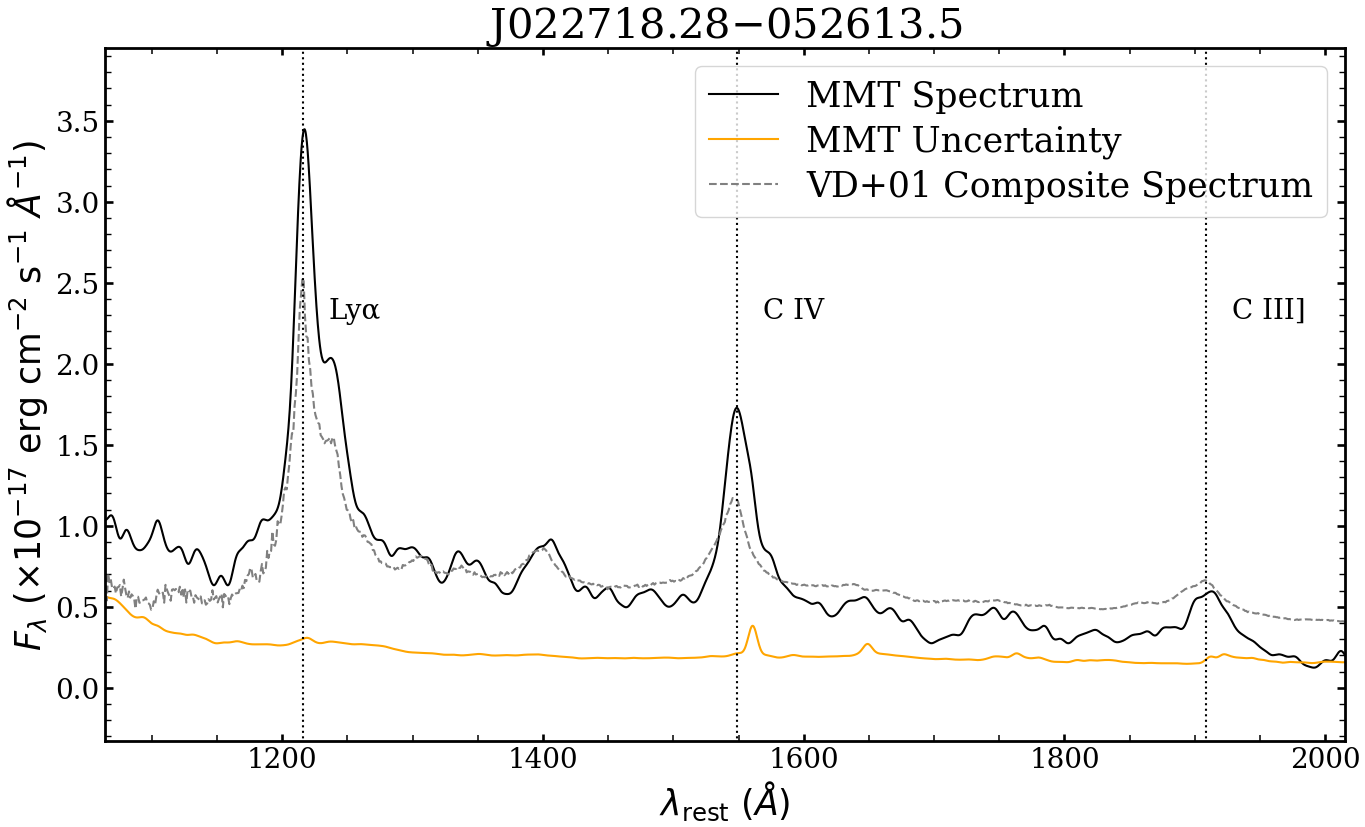}
	\caption{
		The MMT spectrum of J022718.28$-$052613.5 at $z_{\rm spec}=2.573$.
		The black and orange solid curves represent the flux and its uncertainty, respectively.
		We corrected the spectrum for the Galactic extinction.
		We smoothed the spectrum with a Gaussian kernel with a standard deviation of 10 pixels.
		We compared this spectrum to the composite quasar spectrum of \cite{VandenBerk2001}, which is represented as the dashed grey curve and normalized at \hbox{rest-frame} 1450 \AA.
		The photometric redshift predicted by the \hbox{XGBoost} model is $2.52$.
	}
	\label{fig:figvali1}
\end{figure}

\begin{figure}
	\centering
	\includegraphics[scale=0.25]{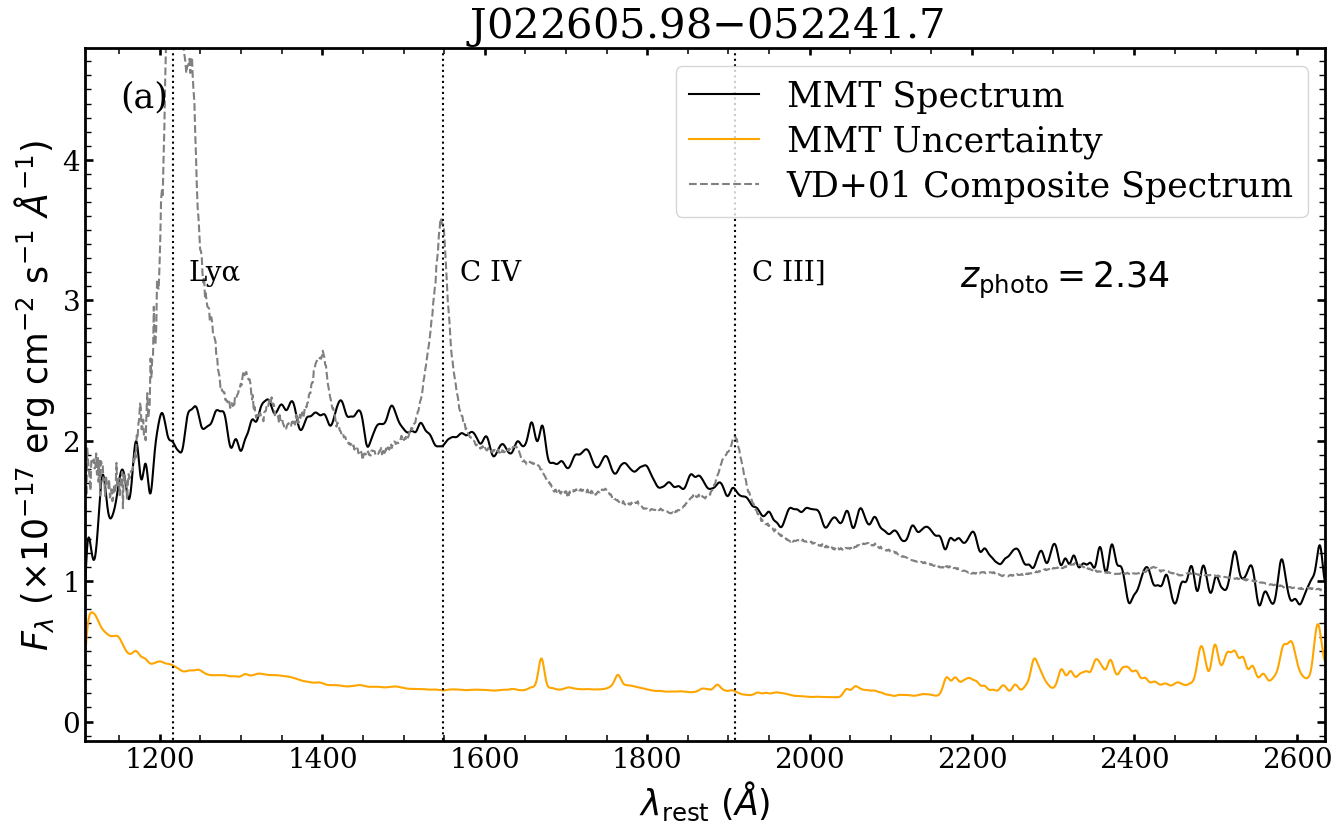}
	\includegraphics[scale=0.25]{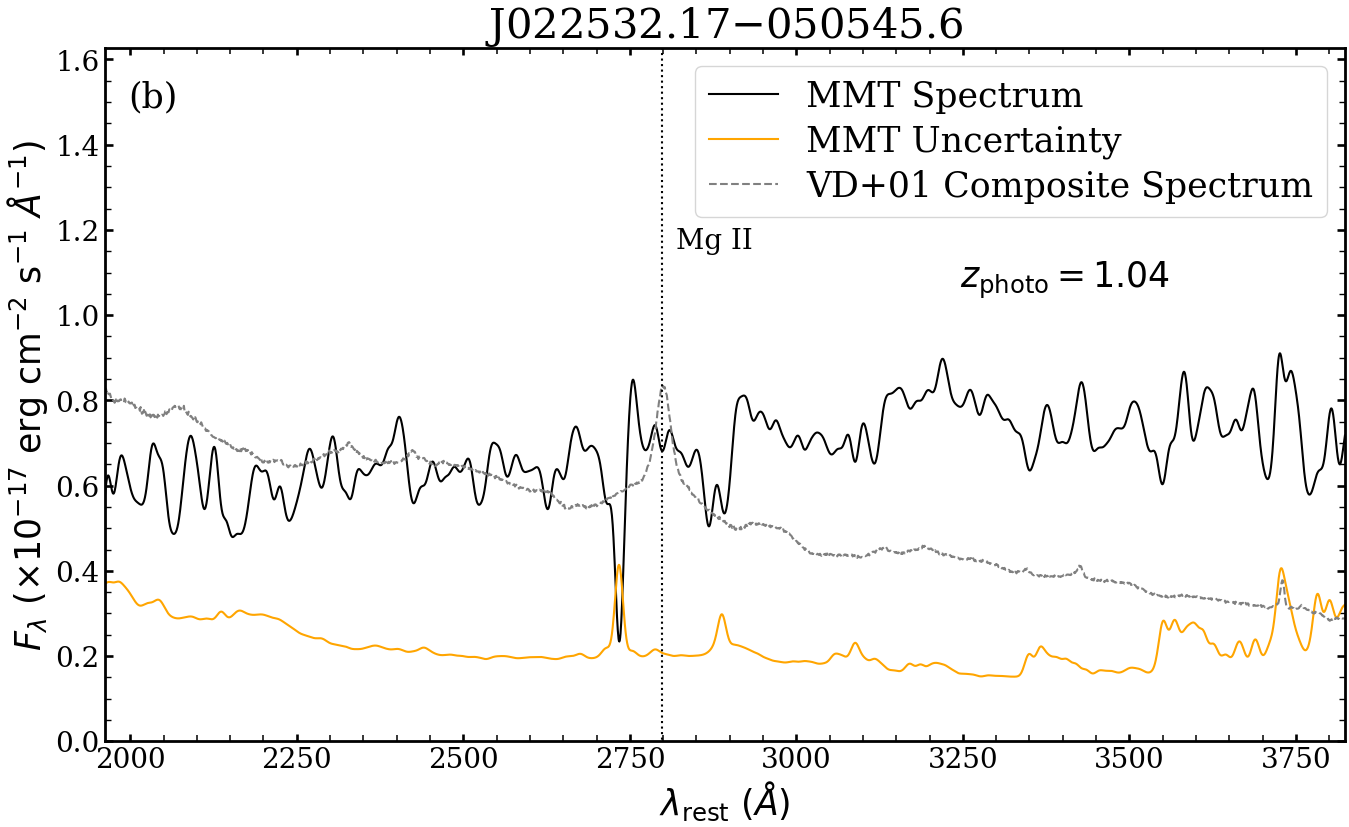}
	\caption{
		The MMT spectra of (a) J022605.98$-$052241.7 and (b) J022532.17$-$050545.6, plotted at their photo-$z$s of 2.34 and 1.04, respectively.
		There are no apparent broad emission lines in the two spectra.
		We compared these two spectra to the composite quasar spectrum from \cite{VandenBerk2001}, which is normalized at \hbox{rest-frame} 2500 \AA.
		These two objects are probably weak \hbox{emission-line} quasars.
	}
	\label{fig:figvali2}
\end{figure}

We visually checked the spectra of the 44 quasars with \hbox{signal-to-noise} ratios $\ge3$ in the observed wavelength range of 4000--7000~{\AA} (5000--7000~{\AA} for WIYN/Hydra spectra), a window where \hbox{telluric-absorption} effect is small.
We identified 41 objects with clear broad emission lines, categorizing them as reliable quasars.
In Figure~\ref{fig:figvali1}, we provide the spectrum of J022718.28$-$052613.5 as an example of such quasars.
We not only successfully classified this quasar but also obtained an accurate estimate of its redshift ($z_{\rm photo}=2.52$ vs. $z_{\rm spec}=2.573$; see Section~\ref{sec:photo_z} below).
We did not find apparent broad emission lines for the other 3 objects, and none of them is \hbox{\xray-detected}.
Among the 3 objects, one is likely a type 2 AGN, with an apparently redder UV/optical continuum compared to the SDSS quasar composite spectrum constructed by \cite{VandenBerk2001}, a clear narrow ${\rm H}\beta$ emission line, and a strong set of [\ion{O}{3}] doublet.
In Figure~\ref{fig:figvali2}, we show the spectra of the other two objects, J022605.98$-$052241.7 and J022532.17$-$050545.6.
There are no apparent broad emission lines in the two MMT spectra.
For J022605.98$-$052241.7, the continuum shape is consistent with that of the SDSS quasar composite spectrum, suggesting that it is probably a weak \hbox{emission-line} quasar (WLQ; e.g., \citealt{Diamond2009, Shemmer2009, Plotkin2010, Laor2011, Wu2012}).
J022532.17$-$050545.6 has a noisy spectrum, and the continuum is apparently redder than the SDSS quasar composite spectrum.
Given its \hbox{photo-$z$} of 1.04, the only strong broad emission line expected in an AGN spectrum is \ion{Mg}{2} $\lambda2799$.
Thus this object could also be a WLQ with weak \ion{Mg}{2} line emission (considering also the uncertain \ion{Mg}{2} location due to the uncertainty from the \hbox{photo-$z$}).
Therefore, 41--43 of the 44 selected quasars have spectroscopic confirmation.

Furthermore, we matched our selected 1\,591 quasars to the SDSS DR16 archival sources that are not included in our training process because they have $\mathtt{zwarning>0}$.
This effort led to 5 additional objects that satisfy the criteria of spectral \hbox{signal-to-noise} ratios $\ge3$ within the observed wavelength range of 4000--7000~{\AA}.
We found clear and discernible broad emission lines in all 5 objects.
These spectroscopic confirmations further validate the reliability of our selected quasars.

\section{Photometric Redshifts} \label{sec:photo_z}
To study the \hbox{disk-corona} connection, we require redshift information for our selected 1\,591 quasars.
While spectroscopic redshifts (spec-$z$s) are ideal, a significant fraction ($\approx40\%$) of them have no spec-$z$s.
Therefore, we estimated photometric redshifts (photo-$z$s) for our selected quasars.

XGBoost offers a regression mode that can be utilized to estimate photo-$z$s.
We constructed the training sample from the 1\,591 quasars in the \hbox{XMM-SERVS} \hbox{XMM-LSS} field, since the coverage of the VIDEO and CFHTLS surveys is more extensive.
Among the 1\,591 quasars, 651 have SDSS $\mathtt{zwarning=0}$ spec-$z$s.
For the 940 selected quasars without SDSS $\mathtt{zwarning=0}$ \hbox{spec-$z$s}, we collected spectroscopic redshifts from the redshift catalogs of other spectroscopic surveys, including the PRIsm \hbox{Multi-Object} Survey (PRIMUS; \citealt{Coil2011}), the 3D-HST Survey \citep{Skelton2014, Momcheva2016} in the XMM-LSS field, the UKIDSS \hbox{Ultra-Deep} Survey (UDSz; \citealt{Bradshaw2013}, \citealt{McLure2013}), the VIMOS Public Extragalactic Redshift Survey (VIPERS; \citealt{Garilli2014}), and the VIMOS VLT Deep Survey (VVDS; \citealt{LeFevre2013}).
Additionally, we used a compilation of publicly available but unpublished spec-$z$s in the UDS field (named as UDS Compilation in this study).\footnote{An overview of the redshift catalog can be found at \url{https://www.nottingham.ac.uk/astronomy/UDS/data/data.html}.}
We matched our quasars to the redshift catalogs of these surveys using a matching radius of 1\arcsec\ and obtained 371 quasars with matched spec-$z$s.
For 71 of the 371 quasars with at least one secure spec-$z$ (determined from multiple reliable spectral features with $\gtrsim95\%$ confidence levels), we adopted their spec-$z$s in order of the VVDS, VIPERS, UDSz, and UDS Compilation catalogs, which are ranked by the spectral resolution at $r$ band \citep{Chen2018}.
Therefore, we obtained 722 ($651+71$) quasars with SDSS spec-$z$s or other secure spec-$z$s, and these quasars are utilized as the training sample for regression analysis.
The spec-$z$s of the 722 quasars span a range of 0.127 to 3.860, with a median value of 1.574.

The regression model was trained using the same 18 colors that were utilized in our classification models (see Section~\ref{subsec:features}).
We \hbox{fine-tuned} the hyperparameters of the regression model through 5\,000 trials with \textsc{optuna}.
The optimal model was selected based on the lowest mean square error (MSE)\footnote{${\rm MSE}={\frac{1}{N}}{\sum_{i=1}^{N}}(z_{\rm spec}-z_{\rm photo})^{2}$.} calculated from a \hbox{five-fold} cross validation of the training sample.
The MSE of the \hbox{five-fold} cross validation of the training sample is 0.134.
We also used two indicators to evaluate the \hbox{photo-z} quality, the outlier fraction ($f_{\rm outlier}$\footnote{$f_{\rm outlier}$ is defined as the fraction of quasars with $|z_{\rm spec}-z_{\rm photo}|/(1+z_{\rm spec})>0.15$.}; e.g., \citealt{Hildebrandt2010}, \citealt{Luo2010}) and the normalized median absolute deviation (NMAD) of $\Delta{z}$ ($\sigma_{\rm NMAD}$\footnote{$\sigma_{\rm NMAD}=1.48\times{\rm median}(\frac{|\Delta z-{\rm median}(\Delta z)|}{1+z_{\rm spec}})$, which is a robust indicator of the photo-z accuracy after the exclusion of outliers.}; e.g., \citealt{Ilbert2006}, \citealt{Brammer2008}).
The $f_{\rm outlier}$ and $\sigma_{\rm NMAD}$ values of the \hbox{five-fold} cross validation of the training sample are $18.4\%$ and $0.054$, respectively.

Subsequently, we used the optimized regression model to estimate the photo-$z$s of the entire set of 1\,591 photometrically selected quasars.
The estimated photometric redshifts span a range of 0.41 to 3.75, with a median value of 1.64.
We note that estimating photometric redshifts for \hbox{type 1} quasars is challenging due to their typically \hbox{power-law-shaped} SEDs in the optical, resulting in colors that are rather independent of redshift \citep[e.g.,][]{Salvato2019}.
To assess the performance of the optimal regressor, we also constructed a \hbox{blind-test} sample including 361 quasars outside the \hbox{XMM-LSS} field that also have SDSS $\mathtt{zwarning=0}$ \hbox{spec-$z$s} or other secure \hbox{spec-$z$s}.
The \hbox{blind-test} sample is not used for training the regression model.

\begin{figure}
\centering
\includegraphics[scale=0.45]{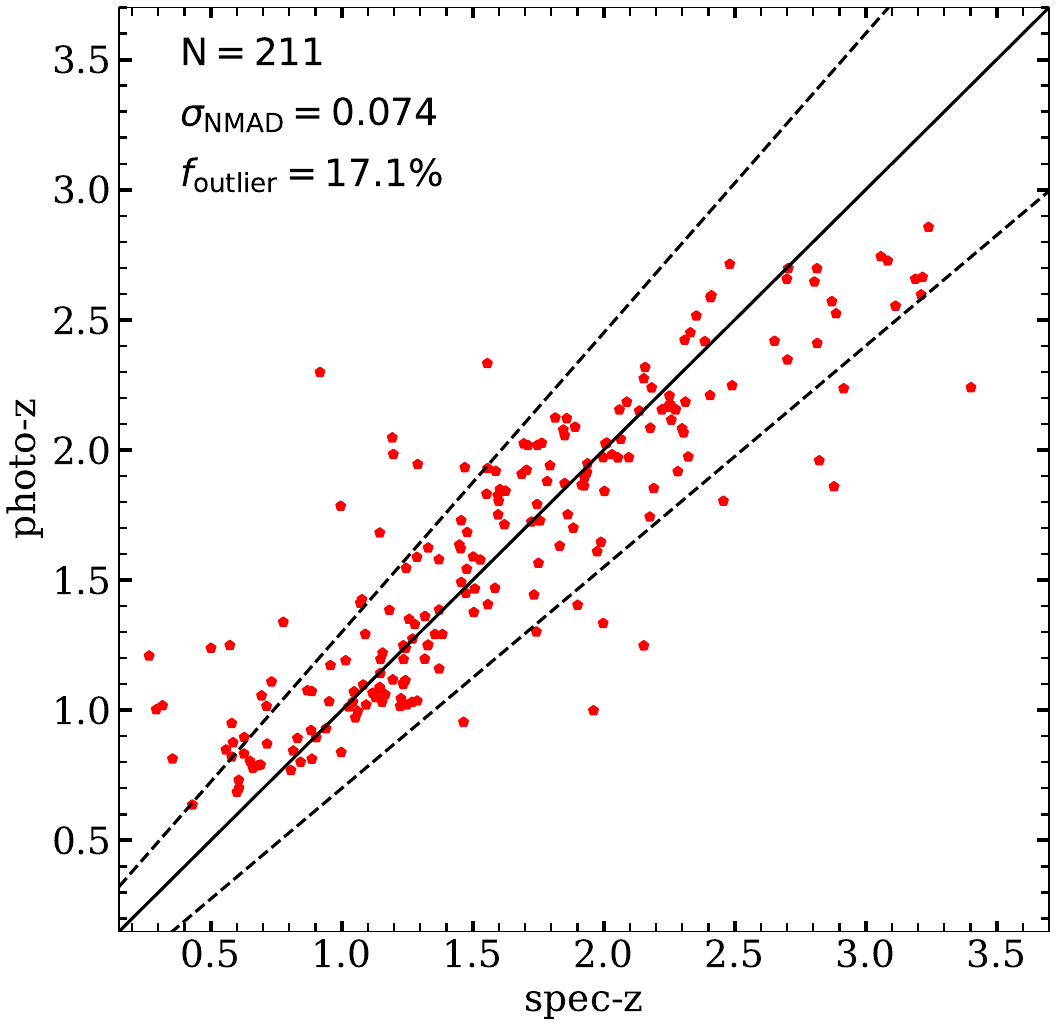}
\caption{
Comparison between the spectroscopic and the photometric redshifts for the 211 quasars with CFHT \hbox{$u^{*}$-band} photometry in the \hbox{blind-test} sample.
The black solid line marks the $z_{\rm spec}=z_{\rm phot}$ relation.
The black dashed lines are the $|{\Delta}z|/(1+z_{\rm spec})=0.15$ thresholds for outliers.
}
\label{fig:fig7}
\end{figure}

We applied the optimal regression model to the \hbox{blind-test} sample.
Upon comparing the photo-$z$s and spec-$z$s of the blind-test sample, we obtained ${\rm MSE}=0.149$, $f_{\rm outlier}=22.7\%$, and $\sigma_{\rm NMAD}=0.079$.
However, approximately 43\% of the \hbox{blind-test-sample} quasars lacked CFHT $u^{*}$ band photometry, whereas the fraction is only around 5\% for the 1\,591 selected quasars in the \hbox{XMM-LSS} field.
The \hbox{$u^{*}$-band} photometry of quasars is powerful for distinguishing between low and high photometric redshift solutions \citep[e.g.,][]{Salvato2019,Sawicki2019} and increasing the accuracy of photometric redshifts \citep[e.g.,][]{Yang2017}.
Therefore, we checked the 211 \hbox{blind-test-sample} quasars that have CFHT $u^{*}$-band photometry, and we found that their photo-$z$s are more reliable, with ${\rm MSE}=0.111$, $f_{\rm outlier}=17.1\%$, and $\sigma_{\rm NMAD}=0.074$.
The comparison between the photo-$z$s and spec-$z$s for these 211 quasars is illustrated in Figure~\ref{fig:fig7}.
For the selected quasars in the XMM-LSS field with no SDSS $\mathtt{zwarning=0}$ spec-$z$s or other secure spec-$z$s, approximately 91\% have CFHT $u^{*}$ band photometry.
Therefore, we estimated $f_{\rm outlier}\approx17\%$ and $\sigma_{\rm NMAD}\approx0.07$ for these selected quasars based on the photo-$z$ accuracy of the \hbox{blind-test-sample} quasars with CFHT \hbox{$u^{*}$-band} photometry.

Our \hbox{machine-learning-based} photo-$z$ quality is comparable to that of previous studies, which reported $f_{\rm outlier}\approx20\%$ (e.g., \citealt{Ni2021}, \citealt{Li2022}).
For comparison, we also utilized the \hbox{Le PHARE} \hbox{SED-fitting} code \citep{Arnouts1999,Ilbert2006}\footnote{\url{https://www.cfht.hawaii.edu/~arnouts/LEPHARE/lephare.html}.} to estimate the photo-$z$s for all the 1\,083 ($722+361$) quasars with SDSS $\mathtt{zwarning=0}$ \hbox{spec-$z$s} or other secure \hbox{spec-$z$s} in the training sample and \hbox{blind-test} sample.
We utilized the SED library designed by \cite{Salvato2009}, which contains 30 templates constructed for fitting SEDs of \hbox{AGN-dominated} sources.
In our analysis, all the available \hbox{13-band} photometric data for each quasar was utilized during the SED fitting.
We adopted a minimum magnitude error of 0.06 for each band \citep{Luo2010} to reduce the influence of data points with unrealistically small errors. 
The \hbox{photo-$z$} results from \hbox{Le PHARE}, with $f_{\rm outlier}=29.5\%$ and $\sigma_{\rm NMAD}=0.072$, were less accurate than those from \hbox{XGBoost}.

\begin{figure}
\centering
\includegraphics[scale=0.38]{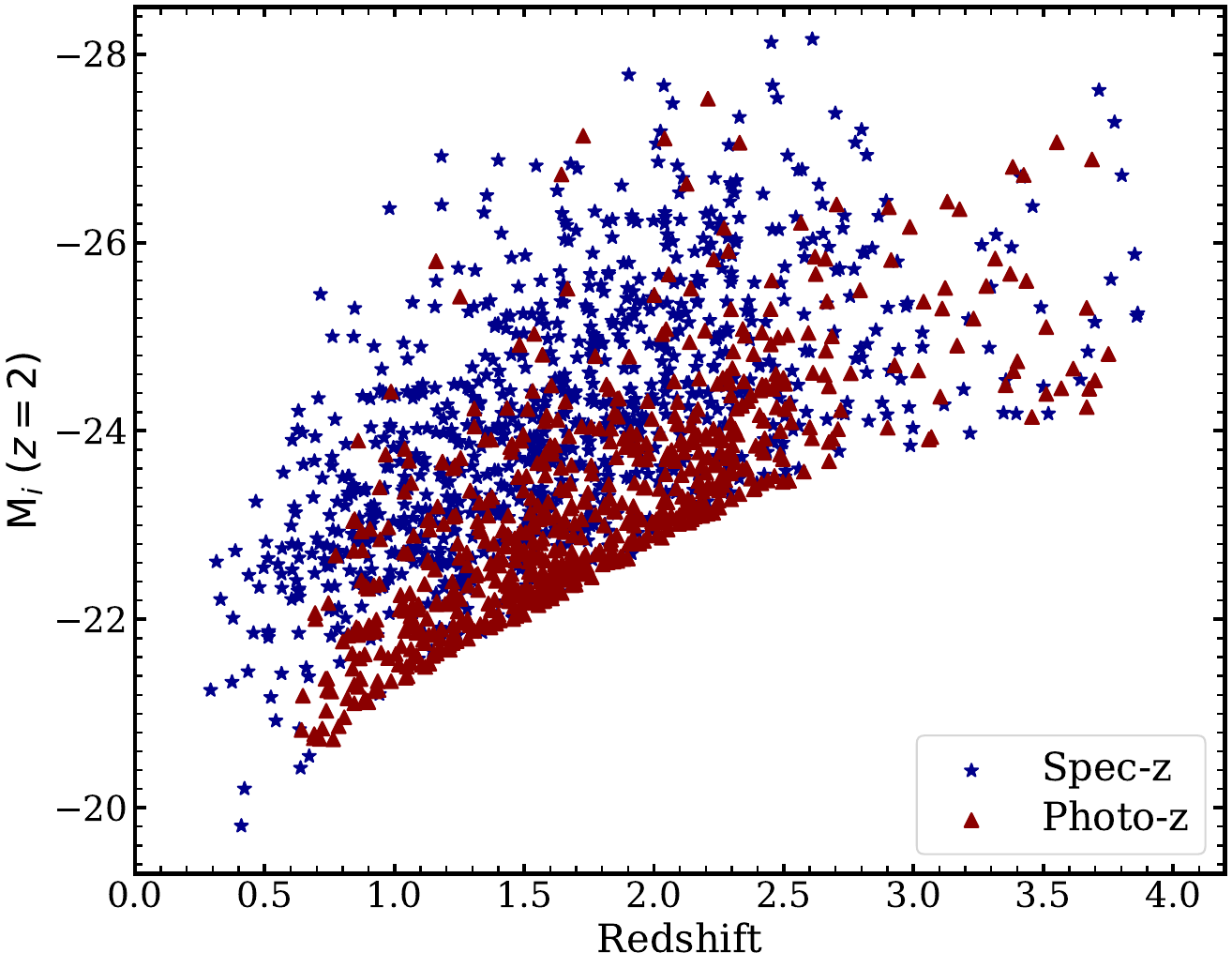}
\caption{
The absolute $i$-band magnitude at $z=2$ vs. redshift for the 1\,591 selected quasars in the \hbox{XMM-SERVS} \hbox{XMM-LSS} field.
The quasars with spec-$z$s or photo-$z$s are represented as the blue stars or the red triangles, respectively.
}
\label{fig:fig6}
\end{figure}

Therefore, we adopt the \hbox{XGBoost} photo-$z$s in the following analyses.
For the 300 quasars with insecure spec-$z$s (including \hbox{spec-$z$s} from PRIMUS and 3D-HST) and 35 quasars with only SDSS $\mathtt{zwarning>0}$ \hbox{spec-$z$s}, we adopted the spec-$z$ if $|z_{\rm spec}-z_{\rm photo}|/(1+z_{\rm spec})<0.15$ (200 objects out of the 335); otherwise, the photo-$z$ is adopted.
Therefore, we have spec-$z$s for 922 ($722+200$) quasars and photo-$z$s for the remaining 669 quasars.
We list the redshifts and the sources of redshift for our selected quasars in Table~\ref{tbl:tbl4}.
In Figure~\ref{fig:fig6}, we display the absolute $i$-band magnitude at $z=2$, $M_i(z=2)$, versus the final adopted redshifts of the selected quasars.
For each quasar, we calculated its $M_i(z=2)$ value using the \hbox{Galactic-extinction-corrected} apparent \hbox{$i$-band} magnitude and the $K$ correction from Table 4 of \cite{Richards2006}.


\begin{deluxetable*}{rrrrrrrrrrrc}[htb!]
\tabletypesize{\scriptsize}
\tablewidth{0pt}
\tablecaption{Optical and \xray\ Properties of Our Selected Quasars}
\tablehead{
\colhead{RA}  &
\colhead{DEC}  &
\colhead{Redshift} &
\colhead{Redshift} &
\colhead{${M}_{i}~(z=2)$} &
\colhead{$f_{2500~ \textup{\AA}}$} &
\colhead{$f_{\rm 2~ keV}$} &
\colhead{$R$} &
\colhead{$\Gamma_{\rm X}$} &
\colhead{$\alpha_{\rm OX}$} &
\colhead{$\Delta \alpha_{\rm OX}$} &
\colhead{$\alpha_{\rm OX}$--${L}_{\rm 2500~\textup{\AA}}$} \\
\colhead{(deg)}  &
\colhead{(deg)}  &
\colhead{ } &
\colhead{Source} &
\colhead{ } &
\colhead{ } &
\colhead{ } &
\colhead{ } &
\colhead{ } &
\colhead{ } &
\colhead{ } &
\colhead{flag}  \\
\colhead{(1)}  &
\colhead{(2)}  &
\colhead{(3)}  &
\colhead{(4)}  &
\colhead{(5)}  &
\colhead{(6)}  &
\colhead{(7)}  &
\colhead{(8)}  &
\colhead{(9)} &
\colhead{(10)} &
\colhead{(11)} &
\colhead{(12)}
}
\startdata
$34.2004$& $-4.9333$& $1.821$& UDSz& $-23.63$& $5.4$& $6.5$& $<5.8$&$1.7$& $-1.50$& $-0.04$& $1$ \\ 
$34.2012$& $-4.4987$& $1.178$& SDSS& $-23.15$& $7.3$& $4.1$& $13.4$&$1.7$& $-1.63$& $-0.20$& $0$ \\ 
$34.2035$& $-5.6902$& $1.932$& VIPERS& $-23.55$& $4.8$& $13.5$& $<73.7$&$1.7$& $-1.36$& $0.11$& $1$ \\ 
$34.2053$& $-3.8765$& $2.08$& XGB& $-24.53$& $10.5$& $12.6$& $<20.1$&$1.7$& $<-1.50$& $<0.02$& $1$ \\ 
$34.2083$& $-5.2951$& $1.280$& UDS& $-22.32$& $3.2$& $6.6$& $<6.1$&$1.64$& $-1.41$& $-0.03$& $1$ \\ 
$34.2092$& $-4.0285$& $1.031$& SDSS& $-23.37$& $11.2$& $27.0$& $<6.2$&$1.13$& $-1.39$& $0.06$& $0$ \\ 
$34.2180$& $-5.6016$& $1.439$& SDSS& $-23.09$& $5.7$& $10.3$& $<24.6$&$1.0$& $-1.44$& $0.01$& $0$ \\ 
$34.2195$& $-5.6841$& $2.612$& SDSS& $-26.04$& $28.2$& $21.6$& $<15.1$&$1.7$& $-1.58$& $0.04$& $1$ \\ 
$34.2206$& $-4.9130$& $1.622$& UDS& $-22.96$& $4.0$& $10.5$& $<4.4$&$1.7$& $<-1.37$& $<0.06$& $1$ \\ 
$34.2233$& $-4.5127$& $1.636$& SDSS& $-24.56$& $17.1$& $5.4$& $<1.6$&$1.7$& $-1.73$& $-0.19$& $1$ \\ 
$34.2282$& $-4.9625$& $1.107$& SDSS& $-23.04$& $9.3$& $19.2$& $<1.5$&$1.35$& $-1.41$& $0.03$& $1$ \\ 
$34.2301$& $-5.1964$& $2.796$& SDSS& $-24.88$& $9.2$& $5.6$& $<2.9$&$1.7$& $-1.62$& $-0.06$& $1$ \\ 
$34.2305$& $-5.3933$& $2.221$& SDSS& $-24.30$& $7.6$& $11.9$& $<3.8$&$1.53$& $-1.46$& $0.05$& $1$ \\ 
$34.2356$& $-5.5001$& $0.69$& XGB& $-20.78$& $1.3$& $8.6$& $266.1$&$1.7$& $<-1.22$& $<0.02$& $0$ \\ 
$34.2398$& $-4.0789$& $1.738$& SDSS& $-23.30$& $5.4$& $14.9$& $<13.2$&$1.79$& $-1.37$& $0.10$& $1$ \\ 
$34.2420$& $-5.5805$& $2.31$& XGB& $-24.84$& $10.8$& $18.7$& $<14.8$&$1.7$& $<-1.44$& $<0.10$& $1$ \\ 
$34.2452$& $-4.2861$& $2.089$& PRIMUS& $-24.07$& $6.9$& $7.3$& $<5.0$&$1.7$& $-1.53$& $-0.02$& $1$ \\ 
$34.2463$& $-4.7591$& $1.071$& SDSS& $-23.08$& $10.8$& $15.4$& $<1.7$&$1.97$& $-1.48$& $-0.03$& $1$ \\ 
$34.2472$& $-5.4082$& $1.426$& SDSS& $-22.90$& $5.2$& $7.4$& $12.6$&$1.7$& $-1.48$& $-0.04$& $0$ \\ 
\enddata
\tablecomments{
Columns (1)-(2): the HSC right ascension and declination (in degrees) of the selected quasar.
Column (3): the redshift.
Column (4): the source of redshift; photo-$z$s are labeled as ``XGB''.
Column (5): the absolute \hbox{$i$-band} magnitude at $z=2$.
Column (6): the \hbox{rest-frame} \hbox{2500$~$\AA} flux density in units of $10^{-29}~{\rm erg}~{\rm cm}^{-2}~{\rm s}^{-1}~{\rm Hz}^{-1}$.
Column (7): the \hbox{rest-frame} \hbox{\rm 2$~$keV} flux density or its $3\sigma$ upper limit in units of $10^{-33}~{\rm erg}~{\rm cm}^{-2}~{\rm s}^{-1}~{\rm Hz}^{-1}$.
Column (8): the radio loudness parameters.
Column (9): the effective photon index, set as ``1.7'' if the quasar is not detected in the hard band (including quasars with no \xray\ detection); or ``1.0'' if the quasar is detected in the hard band but not the soft band.
Column (10): the measured \aox\ parameter.
Column (11): the difference between the measured \aox\ and the expected \aox\ from Equation~\ref{e2}.
Column (12): a flag of ``1'' indicates that the quasar is used for deriving the \hbox{disk-corona} correlation (Equation~\ref{e2}).
The full table contains the optical and \xray\ properties of our 1\,591 quasars selected by \hbox{XGBoost} in the \hbox{XMM-SERVS} \hbox{XMM-LSS} field.
Only a portion of this table is shown here to demonstrate its form and content.
\\
(This table is available in its entirety in machine-readble form.)
}
\label{tbl:tbl4}
\end{deluxetable*}

\begin{deluxetable}{cc}\label{tbl:subsample}
\tablewidth{0pt}
\tablecaption{Step-by-step Construction of the Subsample}
\tablehead{
\colhead{Selection Criteria}     &
\colhead{Number of Quasars}
}
\startdata
In parent sample, &  \\
selected by \hbox{XGBoost} & 2784 \\
\hline
In XMM-LSS field & 1591 (Table~\ref{tbl:tbl4}) \\
\hline
Excluding 150 RL quasars & 1441 \\
\hline
$i<22.5$ & 1132 \\
\hline
Excluding 116  & 1016 (Figure~\ref{fig:aox_l2k5}) \\
X-ray-absorbed quasars  & (including 275 X-ray-undetected) \\
\enddata
\end{deluxetable}

\section{Disk-Corona Connection} \label{sec:disk_corona}
\subsection{Exclusion of Radio-Loud Quasars} \label{subsec:radio_loud}

Radio-loud (RL) quasars may exhibit additional \xray\ emission associated with jets or enhanced coronae (e.g., \citealt{Miller2011, Zhu2020}).
We excluded RL quasars in our analysis of the \hbox{disk-corona} connection.
The \hbox{XMM-LSS} field is fully covered by the Karl G. Jansky Very Large Array (VLA) survey \citep{Heywood2020}, and a fraction ($\gtrsim50\%$) of it is covered by the MeerKAT International Gigahertz Tiered Extragalactic Explorations (MIGHTEE) survey (\citealt{Heywood2022}).
We matched the 1\,591 quasars to the \cite{Zhu2023} source catalog using a matching radius of 1\arcsec, which includes optical/IR sources in the \hbox{XMM-SERVS} fields with the MIGHTEE or VLA counterparts.


We identified 372 quasars with matched radio counterparts.
For each of these quasars, we calculated the \hbox{rest-frame} 6 cm flux density ($f_{\rm 6 cm}$) using the peak flux density of the corresponding radio counterpart and a \hbox{power-law} spectral slope of $a_{\rm r}=-0.8$ (\citealt{Falcke1996}, \citealt{Barvainis2005}).
In cases where a quasar lacks an associated radio counterpart, we estimated an upper limit on its radio flux density, which was set as 5 times the rms noise ($5\sigma_{rms}$) at the source position \citep{Heywood2020,Heywood2022}.
For each quasar, we determined the \ftkf\ value and its associated \ltkf\ value by interpolating or extrapolating its photometric data.
Using these \ftkf\ values and the $f_{\rm 6 cm}$ values for each quasar, we computed the \hbox{radio-loudness} parameter (or its upper limit) as $R = f_{\rm 6 cm}/f_{\rm 2500\textup{\AA}}$ (e.g., \citealt{Jiang2007}).
The $R$ values (or their upper limits) for our quasars are detailed in Table~\ref{tbl:tbl4}.
There are 123 quasars with $R$ measurements exceeding 10, and they were classified as RL quasars \citep[e.g.,][]{Kellermann1989, Jiang2007}.
Besides, 27 quasars with no radio counterparts were found to have $R$ upper limits beyond 100, and we consider them as RL quasar candidates in this study.
All these 150 (123+27) quasars were excluded from the analysis of the \hbox{\aox--\ltkf} relation below.
We still derived the \xray\ properties of these 150 quasars following the approach in Section~\ref{subsec:subsample} below, and these are also listed in Table~\ref{tbl:tbl4}.

\subsection{X-ray Properties and Subsample Construction} \label{subsec:subsample}

\begin{figure}
\centering
\includegraphics[scale=0.45]{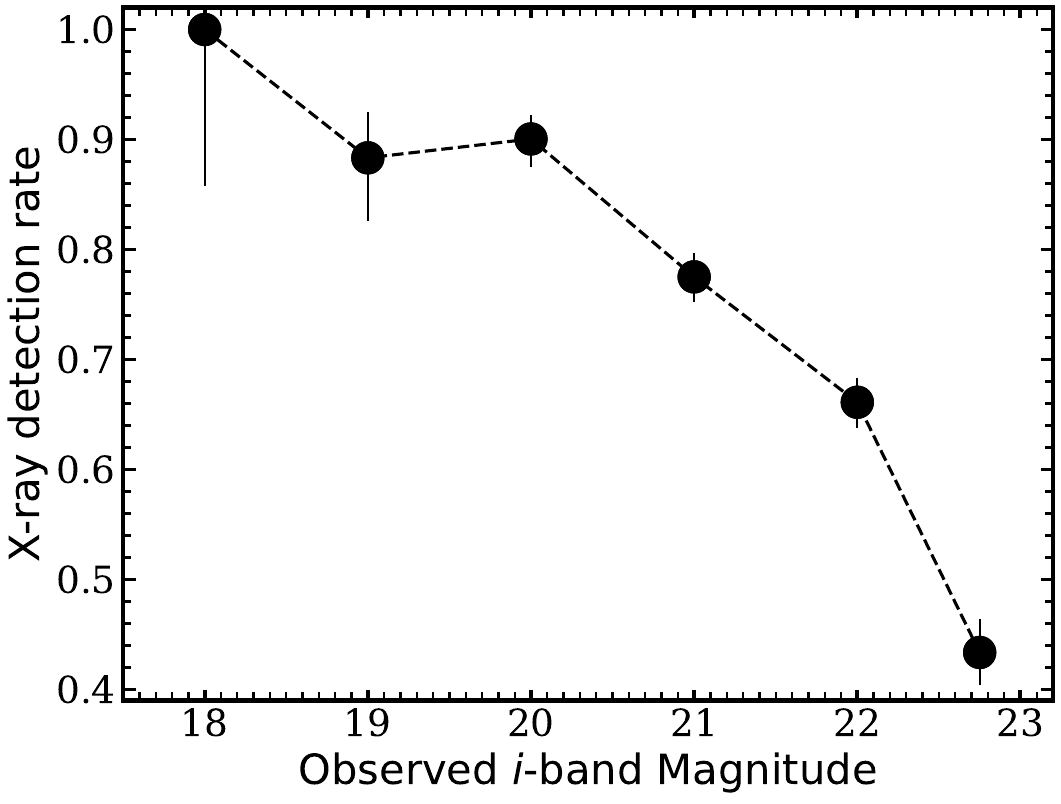}
\caption{
The \xray\ detection rate vs. observed $i$-band magnitude of the 1\,441 selected quasars.
Data are binned with a step of 1 magnitude, spanning the range from 17.5 to 22.5.
The errors are $1\sigma$ statistical uncertainties.
The detection rate drops as the magnitude increases.
}
\label{fig:xdetrate}
\end{figure}

\begin{figure}
\centering
\includegraphics[scale=0.38]{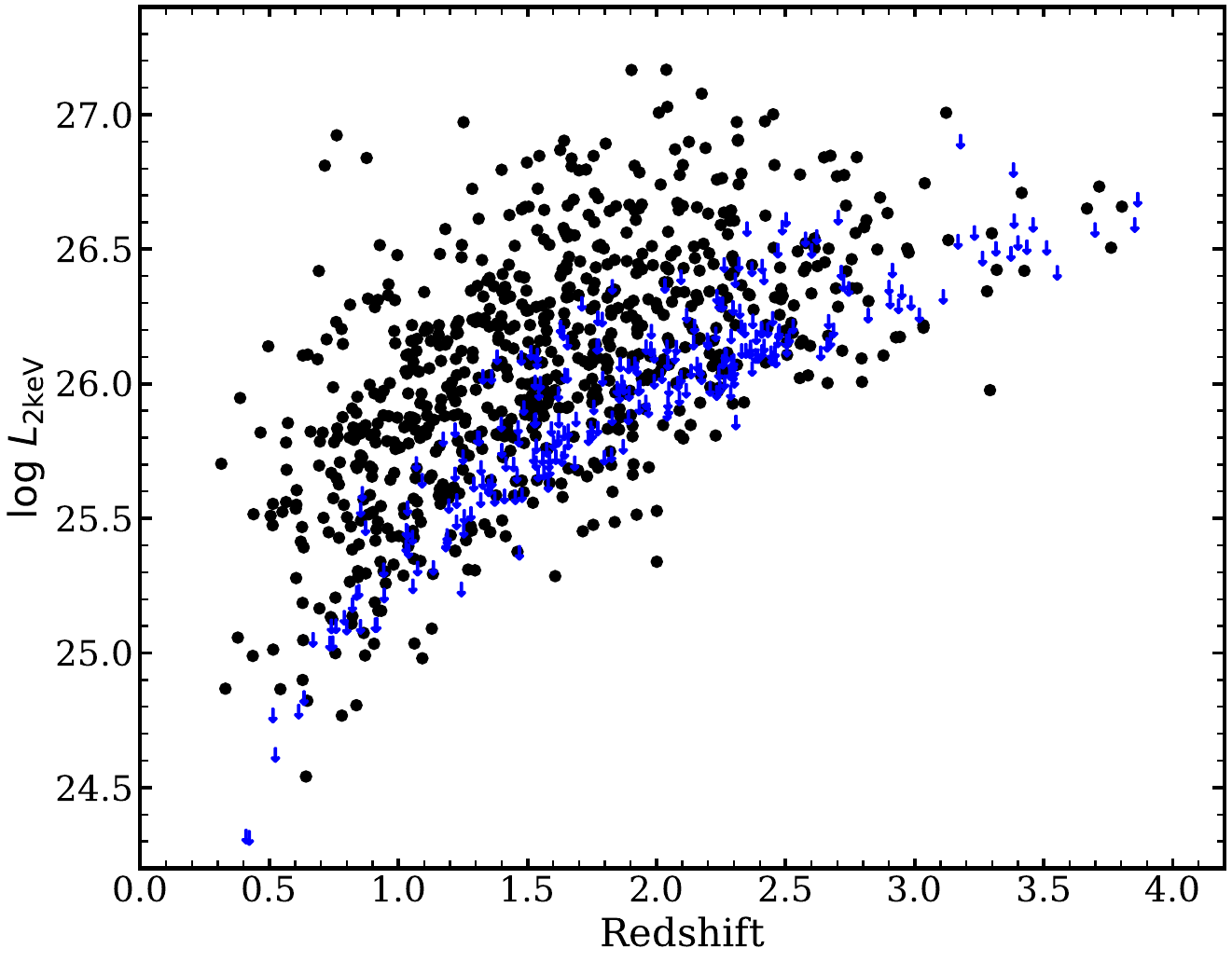}
\caption{
The distribution of \ltkev\ vs. redshifts for the 1\,016 subsample of quasars.
The $3\sigma$ \ltkev\ upper limits of those quasars with no \xray\ detection, calculated from the \hbox{soft-band} sensitivity map, are marked as blue arrows.
}
\label{fig:l2keV_z}
\end{figure}

To obtain the \xray\ properties of our selected quasars, we matched the \xray\ sources (using the coordinates of their \hbox{optical-to-NIR} counterparts) in the catalog of \cite{Chen2018} to our 1\,441 \hbox{radio-quiet} (RQ) quasars (using the HSC coordinates) with an \hbox{1\arcsec\ matching} radius.
A quasar is considered \hbox{\xray-detected} if it has a matched \xray\ counterpart.
We obtained 991 matches ($\approx69\%$), and we show the relationship between the \xray\ detection rate and the $i$-band magnitude of the 1\,441 quasars in Figure~\ref{fig:xdetrate}.
The \xray\ detection rate decreases as the objects become optically fainter, likely due to the survey sensitivity limitation.
More than half of the quasars with $22.5<i<23$ are not detected.

We computed the $f_{\rm 2~keV}$ values for the \hbox{\xray-detected} RQ quasars from their count rates and fluxes in the \cite{Chen2018} catalog:
\begin{enumerate}
	\item 
	For the quasars that are detected in both the soft (\hbox{0.5--2 keV}) and hard (\hbox{2--10 keV}) \xray\ bands, we estimated their \xray\ effective \hbox{power-law} photon indices ($\Gamma_{\rm eff}$) from their \hbox{soft-to-hard-band} ratios, calculated using their \hbox{two-band} count rates of the pn camera \citep{struder2001} provided by the \cite{Chen2018} catalog.
	We adopted a typical pn response file and a simple \hbox{power-law} model modified by Galactic absorption ($N_{\rm H}=3.57\times10^{20}~\mathrm{cm}^{-2}$; \citealt{HI4PI2016}) to determine the $\Gamma_{\rm eff}$ values.
	We calculated the $f_{\rm 2~keV}$ values from their \hbox{soft-band} fluxes using their estimated $\Gamma_{\rm eff}$ values.
	\item 
	For the quasars that are detected in the soft band but not the hard band, we assumed $\Gamma_{\rm eff}=1.7$, which is adopted in \cite{Chen2018}.
	We then calculated their $f_{\rm 2~keV}$ values from their \hbox{soft-band} fluxes.
	\item
	For the other quasars that are detected in the hard band but not the soft band (they are all detected in the full band), we assumed $\Gamma_{\rm eff}=1.0$ and calculated the $f_{\rm 2~keV}$ values from their \hbox{full-band} fluxes.
	We caution that this $\Gamma_{\rm eff}$ value is arbitrarily chosen and the actual spectral shape (and subsequently the $f_{\rm 2~keV}$ value) might be off by a large amount.
	These quasars are not included in our study of the \hbox{\aox--\ltkf} relation below as they are considered likely to be \xray\ absorbed.
	Any future citations to these $f_{\rm 2~keV}$ values should be treated with caution.
\end{enumerate}
For the other quasars that are not \hbox{\xray-detected}, we first derived the $3\sigma$ upper limits on their \hbox{soft-band} fluxes from the survey sensitivity using Equation 3 of \cite{Chen2018}.
We then estimated the $3\sigma$ upper limit on $f_{\rm 2~keV}$ by assuming an \xray\ \hbox{power-law} photon index of 1.7.
In Figure~\ref{fig:l2keV_z}, we present the distribution of $L_{\rm 2keV}$ values versus redshifts for our subsample of quasars.
We list the $\Gamma_{\rm eff}$ and $f_{\rm 2keV}$ values of the 1\,441 quasars in Table~\ref{tbl:tbl4}.

To construct a subsample of quasars with a high \xray\ detection rate for the analysis of the \hbox{\aox--\ltkf} relation (Section~\ref{subsec:correlation} below), we first considered only the 1\,132 RQ quasars with $i<22.5$.
Among these, 857 quasars have matched \xray\ counterparts ($\approx75\%$).
We then excluded quasars that are probably \hbox{\xray-absorbed} from our analysis, as the \xray\ emission does not represent the intrinsic coronal emission.
This exclusion is important to mitigate potential biases when studying the connection between the accretion disk and the corona.
We adopted a criterion of $\Gamma_{\rm eff}<1.26$ that was previously adopted by \cite{Pu2020} and excluded 116 quasars.\footnote{We have also explored alternative $\Gamma_{\rm eff}$ thresholds of 1.0, 1.1, 1.2, and 1.4, and we found that all the derived correlations between \aox\ and \ltkf\ are consistent with each other within the 1$\sigma$ uncertainties.}
We caution that among the \xray\ undetected quasars there should also be \xray\ absorbed ones.
This potential bias is discussed in Section~\ref{subsubsec:xrayabs} below.
The final subsample for studying the quasar \hbox{disk-corona} connection contains 1\,016 quasars.
The step-by-step construction of our subsample is listed in Table~\ref{tbl:subsample} for clarity.

\subsection{The \aox--\ltkf\ Correlation} \label{subsec:correlation}

\begin{figure}
\centering
\includegraphics[scale=0.38]{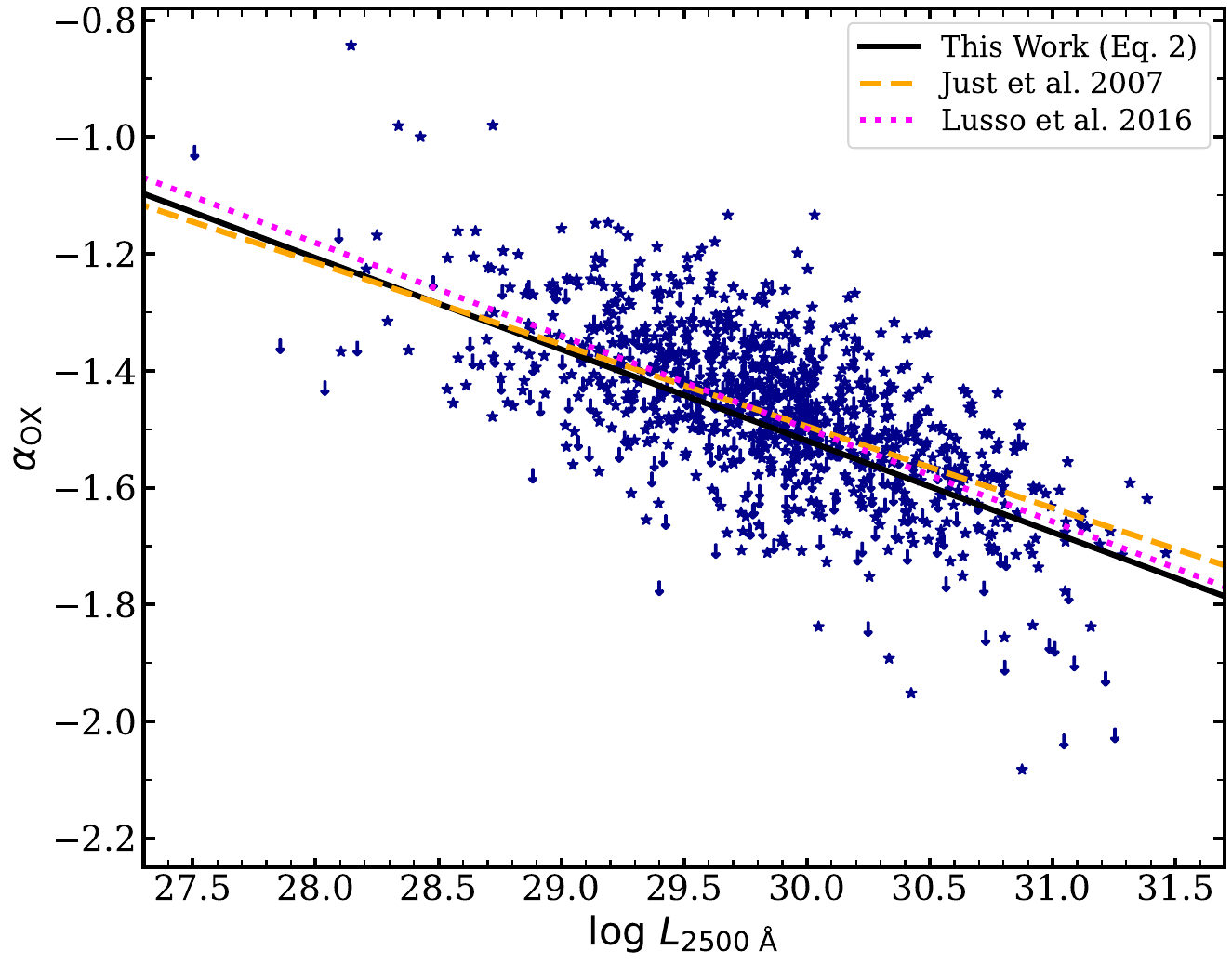}
\caption{
The distribution of \aox\ vs. \ltkf\ values for our 1\,016 subsample of quasars.
Those quasars with no \xray\ detection are marked as arrows.
Our \hbox{\aox--\ltkf} relation is represented by the black solid line, while the \aox--\ltkf\ correlations of \cite{Just2007} and \cite{Lusso2016} are represented as orange dashed line and magenta dotted line for comparison.
Our \hbox{\aox--\ltkf} relation is in good agreement with the two correlations considering the uncertainties.
}
\label{fig:aox_l2k5}
\end{figure}

Using the \ftkev\ values or upper limits in conjunction with the \ftkf\ values, we calculated the \aox\ values or upper limits for our subsample of 1\,016 quasars, including 741 detections and 275 upper limits.
To examine the \hbox{\aox--\ltkf} relation, we utilized the Astronomy Survival Analysis package (ASURV Rev 1.2; \citealt{Isobe1990, LaValley1992}), which is capable of handling censored data.
We confirm a significant \hbox{anti-correlation} (at an $\approx17.7\sigma$ level) between \aox\ and \ltkf\ through the generalized Kendall’s $\tau$ test available in ASURV.
We utilized the EM (estimate and maximize) algorithm in ASURV to determine the linear regression parameters.
In Figure~\ref{fig:aox_l2k5}, we display the derived correlation and the distribution of \aox\ versus \ltkf.
The derived correlation for our subsample of quasars is
\begin{equation}\label{e2}
\begin{split}
\alpha_{\rm OX} & =(-0.156\pm0.007)~{\rm log}~{L_{\rm 2500~\textup{\AA}}} \\
& +(3.175\pm0.211).
\end{split}
\end{equation}
The dispersion of the correlation is calculated as the rms of $\Delta\alpha_{\rm OX}$ \footnote{$\Delta\alpha_{\rm OX}=\alpha_{\rm OX}-\alpha_{\rm OX,~exp}$, where $\alpha_{\rm OX}$ is the measured value, and $\alpha_{\rm OX,~exp}$ is the expected value derived from Equation~\ref{e2}.} obtained from the Kaplan-Meier estimator in ASURV, which is 0.159.
For each quasar, we list its \aox\ and $\Delta\alpha_{\rm OX}$ values in Table~\ref{tbl:tbl4}.

We compared our results to those in two previous studies which also used optically selected AGNs and censored data to study the AGN \hbox{\aox--\ltkf} relation, \cite{Just2007} and \cite{Lusso2016}.
We first compared the numbers of quasars and the log~\ltkf\ and $z$ ranges.
\cite{Just2007} constructed their composite sample, totaling 372 AGNs, by combining the \cite{Strateva2005} AGN sample (most of which are SDSS DR2 AGNs, plus a \hbox{low-redshift} Seyfert 1 sample from \citealt{Walter1993}), photometrically selected COMBO-17 AGNs \citep{Wolf2004}, and a subsample of \hbox{high-luminosity} quasars \citep{Schmidt1983, Vignali2005}.
They obtained the \xray\ properties from several observatories.
The main sample of \cite{Lusso2016} consists of 2\,685 SDSS DR7 quasars from \cite{Shen2011}, with \xray\ properties obtained from the \hbox{3XMM-DR5} source catalog \citep{Rosen2016}.
Our subsample includes a larger number of quasars compared to \cite{Just2007}.
The number of quasars identified could potentially double if our method is applied to the other two \hbox{XMM-SERVS} fields.
The \hbox{low-luminosity} end of the $90\%$ log~\ltkf\ range (from 28.87 to 30.80) of our subsample of quasars is $\approx0.3~{\rm dex}$ higher than that in \cite{Just2007} (28.87 vs. 28.56) and $\approx0.4~{\rm dex}$ lower than that in \cite{Lusso2016} (28.87 vs. 29.31), benefiting from our \hbox{machine-learning} quasar selection approach. 
Therefore, we consider that our method can systematically select a large number of quasars while extending the luminosity range to the \hbox{low-luminosity} end.
We then compared the \hbox{\aox--\ltkf} relations.
Both \cite{Just2007} and \cite{Lusso2016} used ASURV for deriving their \hbox{\aox--\ltkf} relations.
In Figure~\ref{fig:aox_l2k5}, we compare Equation~\ref{e2} to the correlations obtained by \cite{Just2007} and \cite{Lusso2016}.
Notably, Equation~\ref{e2} shows good agreement with the correlations reported in both studies.
We also calculated the correlation dispersion of \cite{Just2007} and \cite{Lusso2016} by using the Kaplan-Meier estimator within ASURV, which are 0.173 and 0.207, respectively.
The dispersion (0.159) of our correlation is slightly smaller than those of \cite{Just2007} and \cite{Lusso2016}.

We also compared our result to that in \cite{Lusso2020}, whose sample contains 2\,421 optically selected quasars with spectroscopic redshifts and \xray\ detections from either Chandra or XMM–Newton.
\cite{Lusso2020} selected quasars that are neither reddened by dust in the optical/UV nor obscured by gas in the \xray, and the \xray\ fluxes are free from flux-limit related biases.
Our subsample has a \hbox{low-luminosity} end of the $90\%$ log~\ltkf\ range $\approx0.3~{\rm dex}$ lower than that in \cite{Lusso2020} (28.87 vs. 29.22).
\cite{Lusso2020} did not derive the \hbox{\aox--\ltkf} relation in their study.
Using their data, we used ASURV to obtain the \hbox{\aox--\ltkf} relation for the 2\,421 quasars.
The relation slope and intercept are $-0.154\pm0.003$ and $3.139\pm0.098$, respectively, and these are in good agreement with those of Equation~\ref{e2} within the $1\sigma$ uncertainties.
The relation has a significantly smaller dispersion of 0.098 compared to Equation~\ref{e2} and those of \cite{Just2007} and \cite{Lusso2016}, which is likely benefiting from using only \hbox{\xray-detected} quasars (see~\ref{subsubsec:xrayabs}). 

\subsection{Potential Effects that May Bias the \hbox{\aox--\ltkf} Relation} \label{subsec:biases}

\subsubsection{Effects from Potentially \hbox{\xray-absorbed} Quasars} \label{subsubsec:xrayabs}
We note that 275 quasars in our subsample are not \xray\ detected.
Given the good sensitivity of the \hbox{XMM-SERVS} survey compared to the brightness cut of our sample construction (Point 2 of Section~\ref{subsec:parent}), it is unlikely that the nondetections were mainly caused by limited \xray\ sensitivity.
Instead, a substantial fraction of these \xray\ undetected quasars should exhibit weak \xray\ emission.
For instance, assuming $\Gamma_{\rm eff}=1.7$ and the $f_{\rm 2~keV}$ values expected from Equation~\ref{e2}, we estimated that 253 (76) of these \xray\ undetected quasars should have \hbox{soft-band} fluxes exceeding the \hbox{soft-band} limiting fluxes (three times the limiting fluxes) if they produce a nominal amount of \xray\ emission ($\Delta\alpha_{\rm OX}=0$).\footnote{The \xray\ weakness of an undetected quasar is generally not reflected by its $\Delta\alpha_{\rm OX}$ upper limit as we used stringent $3\sigma$ upper limits.
For example, for a quasar having expected $f_{\rm 2~keV}$ three times above the sensitivity limit, its $\Delta\alpha_{\rm OX}$ upper limit is $\approx-0.05$.}
Such \xray\ weak quasars might include broad absorption line (BAL) quasars \citep[e.g.,][]{Gallagher2002,Gallagher2006, Fan2009, Gibson2009},  \hbox{weak-line} quasars \citep[e.g.,][]{Luo2015, Ni2018, Ni2022, Timlin2020}, and red quasars \citep[e.g.,][]{Hall2006, Pu2020}, and the \xray\ weakness in these quasars is often ascribed to \xray\ absorption.
Therefore, besides the excluded quasars with $\Gamma_{\rm eff}<1.26$ (see Section~\ref{subsec:subsample}), some of these undetected quasars are probably also \hbox{\xray-absorbed}.
The inclusion of these potentially \hbox{\xray-absorbed} quasars might bias the \hbox{\aox--\ltkf} relation.

We thus performed an analysis of the \hbox{\aox--\ltkf} relation using only the 741 \hbox{X-ray-detected} quasars.
In this subset, we found a correlation slope of $0.157\pm{0.007}$, a correlation intercept of $3.237\pm{0.198}$, and a dispersion of 0.109.
While the new correlation intercept is larger than that of Equation~\ref{e2}, both the new correlation slope and intercept are consistent with those in Equation~\ref{e2} within the $1\sigma$ uncertainties.
This suggests that these potentially \hbox{\xray-absorbed} quasars do not significantly affect the shape of the \hbox{\aox--\ltkf} relation \citep[e.g.,][]{Lusso2016,Lusso2020,Timlin2020}.
Moreover, the dispersion is reduced compared to Equation~\ref{e2}, which is also found in previous studies \citep[e.g.,][]{Lusso2016}.

\subsubsection{Effects from \hbox{Host-galaxy} Contamination and Dust Extinction} \label{subsubsec:redden}
We note that there are no constraints on the estimation of the \ltkf\ values of our subsample of quasars.
However, \hbox{host-galaxy} contamination and dust extinction may both contribute to the reddening of the optical/UV continua of quasars, resulting in overestimation or underestimation of the \ltkf\ values, respectively.
To assess the reddening effects, we tested to isolate a sample of quasars with little UV reddening.

For each quasar in our subsample, we calculated its observed UV \hbox{power-law} spectral slope at \hbox{rest-frame} wavelength of 2500 \AA, denoted as $\alpha_{\nu}$ ($f_{\nu}\propto{\nu}^{\alpha_{\nu}}$).
To derive this spectral slope, we utilized interpolation or extrapolation based on the closest photometric measurements to the quasar's \hbox{rest-frame} 2500 \AA.
We then computed the $\alpha_{\nu,~2500\textup{\AA}}$ value for a reddened quasar SED template.
Specifically, we used composite \#29 in \cite{Salvato2009}, and we modeled the reddened template SED by utilizing the extinction curve of the Small Magellanic Cloud (SMC; \citealt{Gordon2003}) with an extinction value of $E(B-V)=0.15$.
Such an $E(B-V)$ value indicates a change of $\approx0.1$ in the $\Delta\alpha_{\rm OX}$ value.
We then convolved the reddened SED to flux densities with the filter transmission curve corresponding to the two bands we used to derive the observed $\alpha_{\nu,~2500~\textup{\AA}}$ value.
From the flux densities, we calculated the $\alpha_{\nu,~2500~\textup{\AA}}$ value of the reddened template SED.

Out of the 1\,016 quasars in our subsample, we found that 832 quasars have observed $\alpha_{\nu,~2500~\textup{\AA}}$ values that are larger than that calculated from the reddened template SED.
These 832 quasars are likely not significantly affected by \hbox{host-galaxy} contamination or dust extinction.
The average \hbox{log~\ltkf} of the other 184 quasars is 29.4, which is 0.5 dex lower than that of the 832 quasars (29.9).
We analyzed the \hbox{\aox--\ltkf} relation for these 832 quasars, and we found that the result closely resembles Equation~\ref{e2}, with a correlation slope of $-0.155\pm0.008$, an intercept of $3.124\pm0.246$, and a dispersion of 0.144.
This result is consistent with findings from previous studies \citep[e.g.,][]{Lusso2010, Lusso2016, Chiaraluce2018}, which suggest that \hbox{host-galaxy} contamination and dust extinction do not have significant effects on the derived \hbox{\aox--\ltkf} relation for \hbox{type 1} quasars.
Additionally, the slightly smaller dispersion (0.144) compared to that (0.159) of Equation~\ref{e2} suggests that the exclusion of quasars affected by significant UV reddening has the potential to reduce the dispersion of \hbox{\aox--\ltkf} relation \citep[e.g.,][]{Chiaraluce2018}.

\subsubsection{Effects from Extreme X-ray Variability} \label{subsubsec:variability}


Previous studies suggested that \hbox{non-simultaneous} UV and \xray\ measurements has limited effects on the scatter of the \hbox{\aox--\ltkf} relation, with the intrinsic variability of \aox\ values of individual quasars contributing $\approx40-60\%$ of the dispersion \citep[e.g.,][]{Vagnetti2010, Vagnetti2013, Chiaraluce2018}.
The typical \xray\ flux variability amplitudes for RQ \hbox{type 1} quasars range from $\approx20\%$ to $50\%$, which is generally ascribed to instability/fluctuation of the accretion disk and corona \citep[e.g.,][]{McHardy2006,MacLeod2010,Yang2016, Zheng2017}.
Besides the stochastic coronal variability, a number of type 1, RQ, and \hbox{non-BAL} quasars have been found to exhibit extreme and sometimes rapid (down to hours) \xray\ variability with large variation amplitudes (factors of $\gtrsim10$; e.g., PHL 1092, \citealt{Miniutti2012}; SDSS J075101.42$+$291419.1, \citealt{Liu2019}; SDSS J153913.47$+$395423.4, \citealt{Ni2020}; SDSS J135058.12$+$261855.2, \citealt{Liu2022}; SDSS J081456.10$+$532533.5, \citealt{Huang2023}; SDSS
J142339.87$+$042041.1, \citealt{Zhang2023}; SDSS J152156.48$+$520238.5, \citealt{Wang2024}).
Their optical/UV and IR variability during the same period remains limited, suggesting no significant changes in accretion rates.
One possible explanation for such variability is varying obscuration from clumpy \hbox{accretion-disk} winds \citep[e.g.,][]{Liu2022, Wang2022, Huang2023}.
In this case, only the \hbox{high-state} \xray\ emission of these quasars may represent the unabsorbed, intrinsic coronal emission, and studies of their intrinsic \hbox{disk-corona} connection should be based on \hbox{high-state} \xray\ properties.
For example, \cite{Liu2021} studied the \hbox{\aox--\ltkf} relation for a sample of AGNs by using only \hbox{high-state} \xray\ data, and most of the AGNs have simultaneous UV measurements.
They found a smaller dispersion ($\approx0.07$) of the \hbox{\aox--\ltkf} relation compared to previous studies.

In our study, while the selection of quasars with $\Gamma_{\rm eff}>1.26$ helps  us to obtain the \hbox{high-state} \xray\ properties, sometimes the \hbox{low-state} \xray\ spectra dominated by leaked \xray\ emission may also show large $\Gamma$ values \citep[e.g.,][]{Liu2019, Wang2022, Huang2023}.
Therefore, we checked whether there are such quasars with extreme \xray\ variability in our subsample.
Based on the analysis of the \xray\ variability of \hbox{4XMM-DR13} catalog \citep{Webb2020, Traulsen2020} sources in the \hbox{XMM-SERVS} \hbox{XMM-LSS} field, Zhang Z.J. et al (submitted) discovered 8 AGNs with \xray\ variation amplitude ranging from 6 to 12, and only 3 of these are in our subsample.
This result is consistent with the low occurance rate ($\lesssim1-4~\%$) of quasar extreme \xray\ variability \citep[e.g.,][]{Gibson2012,Timlin2020b}.
Therefore, the limited number of quasars in our subsample exhibiting extreme X-ray variability suggests that their effects on our \hbox{\aox--\ltkf} relation is likely not significant.

\subsection{Evolution with $L_{\rm 2500~\textup{\AA}}$ or $z$?} \label{subsec:evolution}
Previous studies have suggested possible evolution of the \hbox{\aox--\ltkf} relation with respect to \ltkf\ or redshift, as the relation slope for \hbox{high-\ltkf} or \hbox{high-$z$} quasars appears steeper \citep[e.g.,][]{Steffen2006, Gibson2008, Chiaraluce2018, Pu2020, Timlin2020}.
For example, \cite{Pu2020} studied the \hbox{\aox--\ltkf} relation for a sample of \hbox{high-\ltkf} SDSS quasars with a $90\%$ log~\ltkf\ range from 29.9 to 31.2, and they found a much steeper relation slope of $-0.224\pm0.008$, which differs from Equation~\ref{e2} ($-0.156\pm0.007$) at an $\approx6.7\sigma$ level.
To investigate whether such an evolution is present in our subsample of quasars, we selected a subset of 211 \hbox{high-\ltkf} SDSS quasars from our subsample by adopting the same redshift ($1.7<z<2.7$) criterion that was used in \cite{Pu2020}.
The $90\%$ log~\ltkf\ range of these 211 quasars is from 29.8 to 31.0, comparable to that of \cite{Pu2020}.
We then used ASURV to derive the \hbox{\aox--\ltkf} relation for the 212 quasars.
The relation slope and intercept are $-0.191\pm0.020$ and $4.256\pm0.594$, respectively.
Despite the significance of difference being only at an $\approx1.7\sigma$ level, the new relation slope is steeper compared to that of Equation~\ref{e2}.
The significance of this trend for our quasars is low due to the limited number of $1.7<z<2.7$ SDSS quasars.
We also divided all of our subsample of quasars into two groups using $z=1.7$ as a threshold.
The median values of \ltkf\ for the $z<1.7$ (\hbox{low-\ltkf}) and $z>1.7$ (\hbox{high-\ltkf}) groups are 29.5 and 30.1, respectively.
The \hbox{\aox--\ltkf} relation slopes for the two groups are $-0.144\pm0.011$ and $-0.182\pm0.013$, respectively.
The evolutionary trend still exists as the two relation slopes differ at an $\approx2.2\sigma$ level.
We note it is difficult to disentangle the degeneracy between \ltkf\ and $z$ for a flux limited quasar sample, and thus the evolution may be related to either of these.

We further tested this possible evolution of the \hbox{\aox--\ltkf} relation with respect to \ltkf/$z$ using the quasars from \cite{Lusso2016} and \cite{Lusso2020}.
We selected the $1.7<z<2.7$ quasars from \cite{Lusso2016} and \cite{Lusso2020} and derived the corresponding \hbox{\aox--\ltkf} relations using ASURV.
The new relation slopes for \cite{Lusso2016} and \cite{Lusso2020} are $-0.228\pm0.013$ and $-0.181\pm0.007$, respectively.
Compared to the relation slopes for their full samples ($0.159\pm0.005$ and $0.154\pm0.003$), the differences are at $\approx4.9\sigma$ and $\approx3.2\sigma$ levels, respectively.

Therefore, our analyses suggest that the \hbox{\aox--\ltkf} relation exhibits an evolution where the correlation becomes steeper in the \hbox{high-\ltkf}/\hbox{high-$z$} regime.
This evolution might be driven by the dependence of \aox\ on more fundamental parameters of SMBH such as $L_{\rm Bol}/L_{\rm Edd}$ and $M_{\rm BH}$ \citep[e.g.,][]{Shemmer2008, Lusso2010, Chiaraluce2018, Liu2021}.
For example, \cite{Liu2021} proposed that $\alpha_{\rm OX}=-0.13{\rm log}(L_{\rm Bol}/L_{\rm Edd})-0.10{\rm log}M_{\rm BH}-0.69$, based on a sample of 47 AGNs with reverberation mapping measured $M_{\rm BH}$ values.
Such different coefficients on the $L_{\rm Bol}/L_{\rm Edd}$ and $M_{\rm BH}$ parameters would lead to different \hbox{\aox--\ltkf} relations for samples with different $L_{\rm Bol}/L_{\rm Edd}$ and $M_{\rm BH}$ distributions.
However, $L_{\rm Bol}$ and $M_{\rm BH}$ estimates for typical quasars have large uncertainties, and it is not even clear that the \aox\ versus $L_{\rm Bol}/L_{\rm Edd}+M_{\rm BH}$  dependence is more fundamental than the \aox\ versus \ltkf\ dependence \citep[e.g.,][]{Chiaraluce2018, Liu2021}.
Therefore, future larger quasar samples with \hbox{well-estimated} $L_{\rm Bol}$ and $M_{\rm BH}$ are needed to reveal the nature of the \hbox{\aox--\ltkf} relation.

\section{Conclusion and Future Work} \label{sec:conclu_future}
In this study, we have selected a sample of quasars located in the \hbox{XMM-LSS} field using a \hbox{machine-learning} approach.
Our primary objective is to investigate the \hbox{disk-corona} connection in these quasars.
We summarize our study as follows:

\begin{enumerate}
\item
We constructed a parent sample comprising 168\,805 sources that were detected in the HSC $g$, $r$, $i$, $z$, and $y$ bands, as well as the Spitzer DeepDrill IRAC $3.6~\micron$ and $4.5~\micron$ bands. 
We then matched the CFHTLS, GALEX, and VIDEO catalog sources to our parent sample to retrieve available photometric data from these surveys.
See Section~\ref{subsec:parent} and Section~\ref{subsec:uv_data}.

\item
We have created both training and \hbox{blind-test} samples by extracting SDSS spectroscopically identified \hbox{type 1} quasars, galaxies, and stars from our parent sample.
The training sample is confined to the \hbox{XMM-LSS} field, while the blind-test sample extends beyond this region.
See Section~\ref{subsec:spec_training} and Section~\ref{subsec:spec_test}.

\item
We utilized the \hbox{XGBoost} algorithm, trained using the training sample, to select a total of 1\,591 quasars from the \hbox{parent-sample} sources located within the \hbox{XMM-SERVS} \hbox{XMM-LSS} field.
We assessed its classification performance based on the \hbox{blind-test} sample.
The outcome was favorable, demonstrating high reliability (approximately $99\%$) and good completeness (approximately $90\%$).
See Section~\ref{sec:selection}.

\item
We used \hbox{XGBoost} to estimate the photometric redshifts of our quasars.
The training sample comprises the photometrically selected quasars with SDSS or other secure spec-$z$s in the \hbox{XMM-LSS} field.
The accuracy of these photometric redshift estimates is acceptable, with $f_{\rm outlier}\approx17\%$ and $\sigma_{\rm NMAD}\approx0.07$.
Furthermore, we gathered spectroscopic redshift data from various survey catalogs for our quasars.
See Section~\ref{sec:photo_z}.

\item
We matched the selected quasars to the \xray\ source catalog of \cite{Chen2018}.
We utilized the measurements available in the \xray\ source catalog to calculate the \xray\ properties of the quasars.
For quasars that were \hbox{X-ray} undetected, we computed the $3\sigma$ upper limit on $f_{\mathrm{2~keV}}$ using the \hbox{soft-band} sensitivity map constructed by \cite{Chen2018}.
See Section~\ref{subsec:subsample}.

\item
We studied the \hbox{\aox--\ltkf} relation for a subsample consisting of 1\,016 RQ quasars where quasars with possible \xray\ absorption were excluded.
See Sections~\ref{subsec:radio_loud} and \ref{subsec:subsample}.
Our derived correlation (Equation~\ref{e2}) has a correlation slope of ($-0.156\pm0.007$), an intercept of ($3.175\pm0.210$), and a dispersion of 0.159.
The correlation is in good agreement with those in \cite{Lusso2016} and \cite{Just2007}.
See Section~\ref{subsec:correlation}.

\item 
We explored several factors that may bias the \hbox{\aox--\ltkf} relation, including potentially \hbox{\xray-absorbed} quasars, UV reddening, and extreme \xray\ variability. 
We found no significant effects from these factors on the derived \hbox{\aox--\ltkf} relation.
See Section~\ref{subsec:biases}.

\item
Consistent with previous studies, we identified an evolutionary trend of the \hbox{\aox--\ltkf} relation for our subsample of quasars, where it becomes steeper in the higher \ltkf/$z$ regime.
This trend is also apparent from our analyses of the quasars in \cite{Lusso2016} and \cite{Lusso2020}. 
See Section~\ref{subsec:evolution}.

\end{enumerate}

Our future studies will focus on the other two \hbox{XMM-SERVS} regions, the $\approx4.6~{\rm deg}^{2}$ W-CDF-S and $\approx3.2~{\rm deg}^{2}$ ELAIS-S1 regions, where the \xmm\ source catalogs are presented in \cite{Ni2021}.
Both of these two regions are covered by HSC and DeepDrill surveys, similar to the \hbox{XMM-LSS} field.
We intend to utilize our quasar selection methods on the photometric sample sources within these two regions to obtain a larger sample of quasars, enabling us to better investigate the connection between the accretion disk and the corona.
Forthcoming spectroscopic data in the \hbox{XMM-SERVS} regions from instruments such as the Prime Focus Spectrograph \citep[e.g.,][]{Greene2022}, \hbox{Multi-Object} Optical and \hbox{Near-infrared} Spectrograph \citep[e.g.,][]{Maiolino2020}, and \hbox{4-metre} \hbox{Multi-Object} Spectroscopic Telescope \citep{Swann2019} will provide a larger training sample and better redshift measurements which should benefit greatly the photometric selection of quasars and study of the \hbox{\aox--\ltkf} relation.

This study may also be applicable to the forthcoming large samples of photometric quasars selected from the \hbox{wide-field} multiband imaging and spectroscopic survey of the Chinese Space Station Telescope (CSST; \citealt{Zhan2011}).
The CSST wide-field multiband imaging survey is aimed to image approximately 17\,500 square degrees of the sky using NUV, $u$, $g$, $r$, $i$, $z$, and $y$ bands, achieving a $5\sigma$ limiting magnitude of 25.4 (AB mag) or higher for point sources in the NUV and u bands.
The CSST wide-field slitless spectroscopic survey will also cover approximately 17\,500 square degrees of the sky using GU, GV, and GI bands, achieving a $5\sigma$ limiting magnitude of 23.2 (AB mag) or higher for point sources in the GU bands.
Combining the CSST imaging survey with future deep IR surveys like Spitzer DeepDrill survey and X-ray surveys like \hbox{XMM-SERVS}, we will be able to select unprecedentedly large samples of quasars, especially \hbox{low-luminosity} quasars, to study their \hbox{disk-corona} connection.
\vspace{5mm}

We thank the anonymous referee for constructive feedback.
We thank Chien-Ting Chen, Yuming Fu, Qi Hao, Linhua Jiang, and Xue-Bing Wu for helpful discussions.
J.H. and B.L. acknowledge financial support from the National Natural Science Foundation of China grant 11991053 and 11991050, China Manned Space Project grants NO. \hbox{CMS-CSST-2021-A05} and NO. \hbox{CMS-CSST-2021-A06}, and CNSA program D050102.
W.N.B. acknowledges support from NSF grants AST-2106990 and AST-2407089 and the Penn State Eberly Endowment.
Y.Q.X. acknowledges financial support from the National Natural Science Foundation of China grants 12025303 and 1239814.
The MMT spectra reported here were obtained at the MMT Observatory, a joint facility of the Smithsonian Institution and the University of Arizona.
The WIYN spectra presented here were based on the observation at the WIYN Observatory, a joint facility of the NSF’s National Optical-Infrared Astronomy Research Laboratory, Indiana University, the University of Wisconsin-Madison, Pennsylvania State University, and Purdue University.
The codes and data supporting the results of this study are publicly available on GitHub\footnote{\url{https://github.com/choubalv-hj/AstroML}.} and have been archived on Zenodo.\footnote{\url{ https://doi.org/10.5281/zenodo.14321895}.}

\software{{\sc astropy} \citep{astropy:2013, astropy:2018, astropy:2022},
          {\sc dustmaps} \citep{2018JOSS....3..695M},
          {\sc optuna} \citep{Akiba2019},
          {\sc scikit-learn} \citep{pedregosa2011scikit},
          TOPCAT \citep{2005ASPC..347...29T},
          \hbox{XGBoost} \citep{Chen2016}.
}

\bibliographystyle{aasjournal}
\bibliography{ms_quasar_selection.bib}

\end{document}